\pdfoutput=1
\RequirePackage{ifpdf}
\ifpdf 
\documentclass[pdftex]{sigma}
\else
\documentclass{sigma}
\fi

\numberwithin{equation}{section}
\usepackage{bbm}

\begin{document}

\newcommand{\arXivNumber}{1403.3038}


\renewcommand{\thefootnote}{$\star$}

\renewcommand{\PaperNumber}{079}

\FirstPageHeading

\ShortArticleName{Group Momentum Space and Hopf Algebra Symmetries in $2+1$ Gravity}

\ArticleName{Group Momentum Space and Hopf Algebra\\
Symmetries of Point Particles Coupled\\
to $\boldsymbol{2+1}$ Gravity\footnote{This paper is a~contribution to the Special Issue on Deformations of Space-Time
and its Symmetries.
The full collection is available at \href{http://www.emis.de/journals/SIGMA/space-time.html}
{http://www.emis.de/journals/SIGMA/space-time.html}}}

\Author{Michele ARZANO~$^\dag$, Danilo LATINI~$^\ddag$ and Matteo LOTITO~$^{\S}$}

\AuthorNameForHeading{M.~Arzano, D.~Latini and M.~Lotito}

\Address{$^\dag$~Dipartimento di Fisica and INFN, ``Sapienza'' University of Rome, P.le A.~Moro 2,\\
\hphantom{$^\dag$}~00185 Roma, Italy}
\EmailD{\href{michele.arzano@roma1.infn.it}{michele.arzano@roma1.infn.it}}

\Address{$^\ddag$~Dipartimento di Fisica and INFN, Universit\`a  di Roma Tre, Via Vasca Navale 84, \\
\hphantom{$^\ddag$}~I-00146 Roma, Italy}
\EmailD{\href{latini@fis.uniroma3.it}{latini@f\/is.uniroma3.it}}

\Address{$^{\S}$~Department of Physics, University of Cincinnati, Cincinnati, Ohio 45221-0011, USA}
\EmailD{\href{lotitomo@mail.uc.edu}{lotitomo@mail.uc.edu}}

\ArticleDates{Received March 13, 2014, in f\/inal form July 15, 2014; Published online July 24, 2014}

\Abstract{We present an in-depth investigation of the ${\rm SL}(2,\mathbb{R})$ momentum space describing point particles
coupled to Einstein gravity in three space-time dimensions.
We introduce dif\/ferent sets of coordinates on the group manifold and discuss their properties under Lorentz
transformations.
In particular we show how a~certain set of coordinates exhibits an upper bound on the energy under deformed Lorentz
boosts which {\it saturate} at the Planck energy.
We discuss how this deformed symmetry framework is generally described by a~quantum deformation of the Poincar\'e group:
the {\it quantum double} of ${\rm SL}(2,\mathbb{R})$.
We then illustrate how the space of functions on the group manifold momentum space has a~dual representation on a~{\it
non-commutative} space of coordinates via a~(quantum) group Fourier transform.
In this context we explore the connection between Weyl maps and dif\/ferent notions of (quantum) group Fourier transform
appeared in the literature in the past years and establish relations between them.}


\Classification{81R50; 83A05; 83C99}

\renewcommand{\thefootnote}{\arabic{footnote}}
\setcounter{footnote}{0}

\section{Introduction}

General relativity in three space-time dimensions of\/fers an unparalleled insight on how quantum group deformed
relativistic symmetries replace ordinary structures linked to the Poincar\'e group.

The simple picture is that relativistic point particles are coupled to the theory as conical defects owing to the
topological nature of Einstein gravity in three dimensions~\cite{Deser:1983tn, STAR}.
For vanishing cosmological constant the resulting space-time is f\/lat everywhere except at the location of the particle,
the conical singularity.
The parametrization of a~moving defect, like e.g.
the conical particle, requires their momentum to be described by an element of the isometry group of the ambient space.
In our specif\/ic case it turns out that momenta are no longer vectors in three dimensional Minkowski space but elements
of the double cover of the Lorentz group in three dimensions, ${\rm SL}(2,\mathbb{R})$~\cite{MATS,Schroers:2007ey}.
Lorentz transformations are easily implemented for this curved momentum space in terms of the action of ${\rm
SL}(2,\mathbb{R})$ on itself.

At the group theoretic level such transition from vector-like to group-like momenta is captured by the transition from
the (group algebra of) the Poincar\'e group to the {\it quantum double} of ${\rm SL}(2,\mathbb{R})$, a~non-trivial Hopf
algebra in which the three-dimensional Newton's constant appears as a~deformation parameter~\cite{Bais:1998yn,bais}.
Functions on ${\rm SL}(2,\mathbb{R})$ provide now the space on which representations of the double of ${\rm
SL}(2,\mathbb{R})$ are def\/ined~\cite{Arzano, Koornwinder:1996uq}.
A~notion of Fourier transform can be introduced which maps these functions on a~group onto functions of {\it
non-commutative space} in which coordinates obey non-trivial commutation relations given by the brackets of the Lie
algebra of ${\rm SL}(2,\mathbb{R})$~\cite{FreidelMajid, Sasai:2009az}.
Alternatively one can map these functions onto ordinary three-dimensional Minkowski space equipped with
a~non-commutative {\it star-product} determined by the non-abelian group structure of ${\rm
SL}(2,\mathbb{R})$~\cite{freidel3d}.
Thus the study of relativistic point particles coupled to three-dimensional gravity leads us to consider all the
important aspects of deformation of relativistic symmetries and non-commutativity within a~clear geometric picture.

Our aim in this work will be to review the main features discussed above and point out some new results which will help
link the structures emerging in this three-dimensional context to other models of deformed symmetries and
non-commutativity in four space-time dimensions.

We start in the next section by reviewing how momenta of conical defects in three dimensions are parametrized by ${\rm
SL}(2,\mathbb{R})$ group elements.
In particular we show how the usual notion of mass-shell describing the momentum of physical particles is replaced~by
the condition that group valued momenta belong to the {\it conjugacy class} of a~rotation by a~def\/icit angle
characterizing the conical defect.
In Section~\ref{section3} we describe coordinate systems on ${\rm SL}(2,\mathbb{R})$ starting with two choices which are popular in
the literature and introducing a~new parametrization based on {\it Euler angles}.
For each parametrization we write down the explicit form of the conjugacy class condition which represents the
realization of a~deformed mass-shell in each choice of coordinates and show that for the new parametrization introduced
on-shell momenta exhibit a~{\it maximum energy}.
Furthermore we discuss the action of Lorentz transformations on the ${\rm SL}(2,\mathbb{R})$ momentum space case~by
case.
In particular we show that the Euler angle coordinates transform non-linearly under the action of a~boost and that the
energy saturates at the maximum value when the boost parameter is let to inf\/inity.
Thus we f\/ind a~set of momentum coordinates which provide an explicit realization of {\it doubly special relativity} for
a~particle coupled to three-dimensional gravity~\cite{AmelinoCamelia:2000mn,amelino}.
In Section~\ref{section4} we review what is the structure of the underlying quantum group symmetry associated with the ${\rm
SL}(2,\mathbb{R})$ momentum space given by the {\it quantum double} of ${\rm SL}(2,\mathbb{R})$ and we show how such
Hopf algebra can be obtained as a~deformation of the group algebra of the Poincar\'e group.
Next we discuss non-commutative plane waves with ${\rm SL}(2,\mathbb{R})$ momenta and the associated notions of
(quantum) group Fourier transform (a thorough investigation on this and a~detailed description of the symmetries arising
in this context can be found in the recent work~\cite{Schroers:2014}).
This gives us the opportunity to clarify the connection between dif\/ferent notions of Fourier transform appearing in the
literature and to discuss their relation with {\it Weyl maps} which have been extensively used in the literature on
non-commutative space-times in $3+1$ dimensions~\cite{Agostini:2003vg,Agostini:2002de,Meljanac:2007xb}.
Finally we provide a~{\it master equation} for the group Fourier transform which for each choice of Weyl map,
corresponding to a~given group parametrization, reproduces the dif\/ferent group Fourier transforms appeared in the
literature.
Section~\ref{section5} is devoted to brief concluding remarks.

\section[The ${\rm SL}(2,\mathbb{R})$ momentum space of particles coupled to $2+1$ dimensional gravity]{The $\boldsymbol{{\rm SL}(2,\mathbb{R})}$ momentum space of particles coupled\\ to $\boldsymbol{(2+1)}$-dimensional gravity}\label{section2}

\subsection[Particles as conical defects in $2+1$ gravity]{Particles as conical defects in $\boldsymbol{2+1}$ gravity}\label{section2.1}

\begin{figure}[htbp]
\centering \includegraphics[width=7cm]{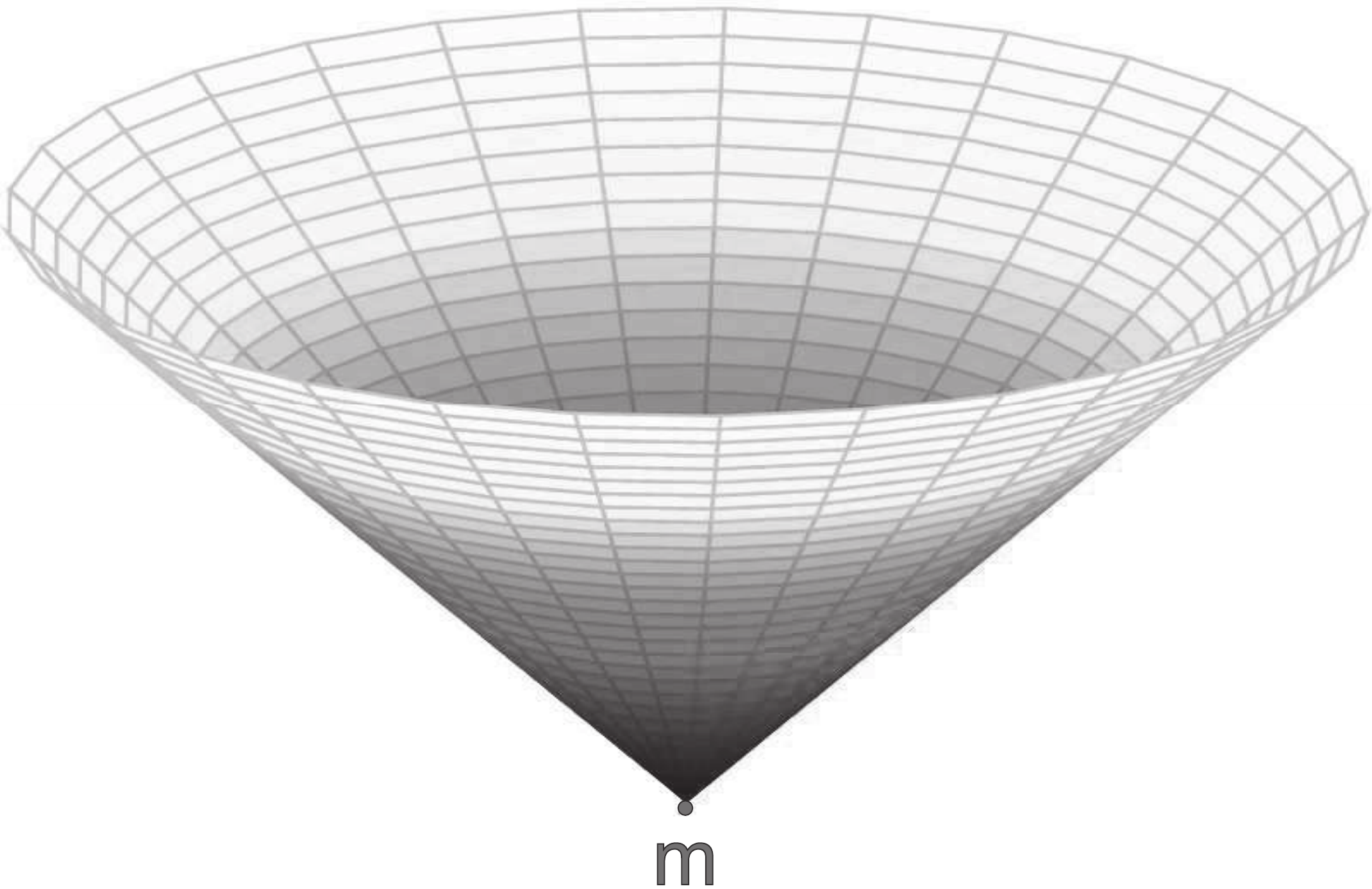}
\caption{Point particle generating a~conical defect.
The particle of mass~$m$ is located at the tip of the cone.}
\label{cone}
\end{figure}

General relativity in three space-time dimensions has the well known property of not possessing local degrees
of freedom.
All solutions of Einstein's equations with vanishing cosmological constant are locally f\/lat and the only non-trivial
degrees of freedom have to be of topological nature.
This must be taken into account when coupling particles to the theory.
Indeed, as already shown by Staruszkiewicz in his seminal work~\cite{STAR}, later thoroughly extended by Deser, Jackiw
and 't~Hooft~\cite{Deser:1983tn}, point-like degrees of freedom carrying mass and spin can be added introducing
``punctures'' on the space-time (f\/lat) manifold.
In the simplest case of a~spinless particle, which will be of interest for the present work, one can picture such
a~puncture as crea\-ting a~{\it conical singularity} resulting in a~space-time which is f\/lat everywhere except at the
location of the particle (see Fig.~\ref{cone}).

This conical space can be characterized by a~{\it deficit angle} $\alpha=8\pi G m$ where~$m$ is the mass of
the particle and~$G$ is Newton's constant in three dimensions.

\begin{figure}[htbp]
\centering \includegraphics[width=11cm]{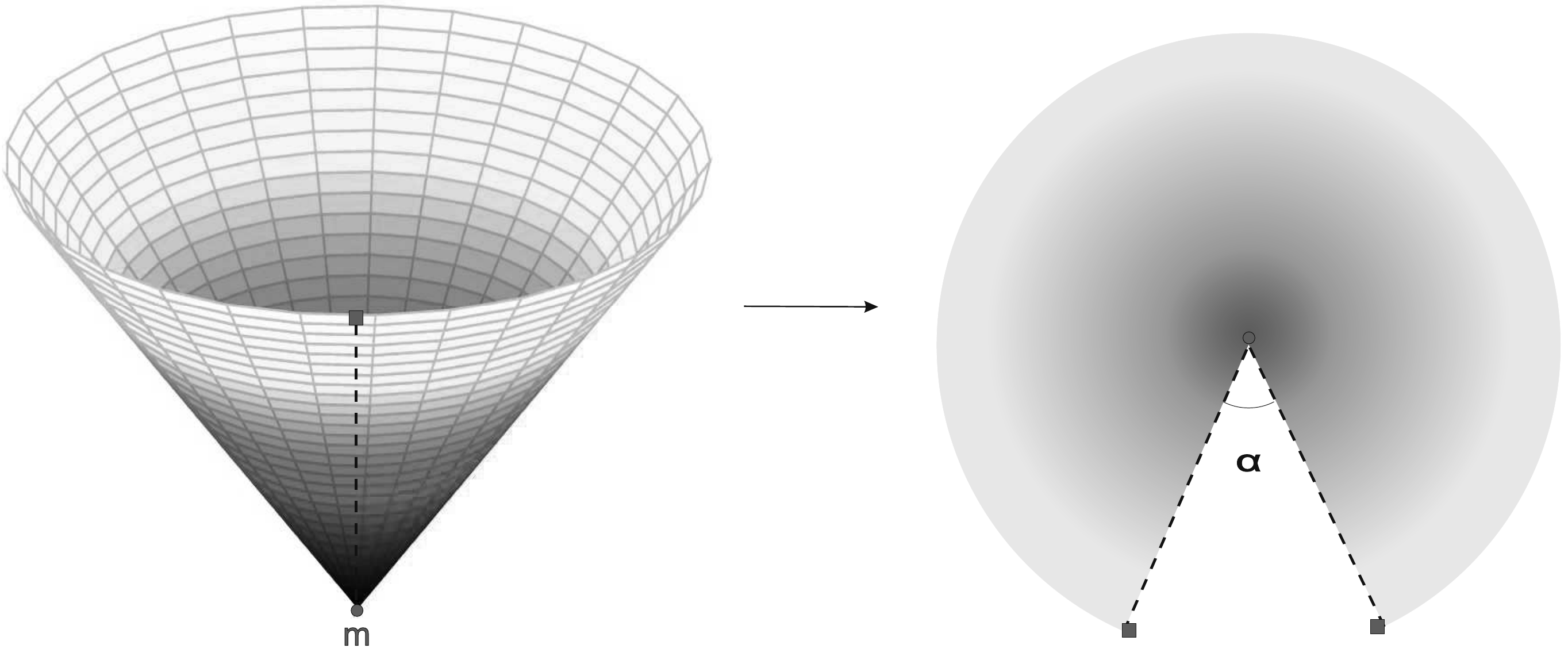}
\caption{The cone is cut out along the dashed
line and f\/lattened on a~plane.
The dashed lines and the squares identify the same points.}
\label{cone2}
\end{figure}

 The def\/icit angle is simply the angle formed by the missing wedge obtained when ``opening up'' the conical
space on a~f\/lat Minkowski space (see Fig.~\ref{cone2}).
The metric of such conical space-time is given~by
\begin{gather*}
ds^2=-dt^2+dr^2+r^2d\varphi^2  ,
\end{gather*}
with the azimuthal angle~$\varphi$ having period $2\pi-\alpha$.
Alternatively we can write the metric keeping the angular~$\varphi$-coordinate with a~period $2\pi$ as:
\begin{gather*}
ds^2=-dt^2+dr^2+(1-4Gm)^2r^2d\varphi^2.
\end{gather*}
The def\/icit angle can be measured by calculating the holonomy of the connection associated to the metric along a~loop
encircling the particle.
This describes the parallel transport of a~vector around the singularity~\cite{MATS}, which results in a~rotation by an
angle $\alpha=8\pi G m$.
Roughly speaking we have that the mass, i.e.~the three-momentum at rest of the point particle/defect, is characterized
by a~rotation proportional to the mass of the particle multiplied by Newton's constant.

To characterize the physical momentum of a~moving defect let us brief\/ly recall how physical momenta of
a~relativistic point particle are described in ordinary Minkowski space.
In three space-time dimensions Minkowski space $\mathbb{R}^{2,1}$ is isomorphic, as a~vector space, to
$\mathfrak{sl}(2,\mathbb{R})$, the algebra of the Lie group ${\rm SL}(2,\mathbb{R})$.\footnote{For our readers
convenience we recall the basic properties of the ${\rm SL}(2,\mathbb{R})$ group in Appendix~\ref{appendixA}.}
If ${\gamma}^{\mu}$ is a~basis of traceless $2 \times 2$ matrices, given an element ${\mathbf p} \in
\mathfrak{sl}(2,\mathbb{R})$ and a~vector $\vec{p} \in \mathbb{R}^{2,1}$ we can write~\cite{MATS}
\begin{gather*}
{\mathbf p}=p^{\mu} \gamma_{\mu}
\quad
\iff
\quad
p^{\mu}=\frac{1}{2}\text{Tr}({\mathbf p} {\gamma}^{\mu})  .
\end{gather*}
Momenta are thus given by the following $\mathfrak{sl}(2,\mathbb{R})$ algebra element
\begin{gather*}
{\mathbf p}=p^{\mu}\gamma_{\mu}=
\begin{bmatrix}
p^2 &  p^1+p^0
\\
p^1-p^0 &  -p^2
\end{bmatrix},
\end{gather*}
whose determinant is $\det {\mathbf p}=(p^0)^2-|\underline{p}|^2$.
We can describe physical momenta by boosting the momentum at rest of a~particle.
The latter will be given in matrix representation by $\overline{{\mathbf p}} = m\gamma_0 =
\begin{bmatrix}
0 & m
\\
-m & 0
\end{bmatrix}
$ and the boost is achieved via the {\it adjoint action} of ${\rm SL}(2,\mathbb{R})$ on $\mathfrak{sl}(2,\mathbb{R})$ so
that a~physical momentum will be given~by
\begin{gather}
{\mathbf p}= {\mathbf h}^{-1} \overline{{\mathbf p}}  {\mathbf h}.
\label{eq:adjoint}
\end{gather}
Such action preserves the determinant and the physical momenta, obtained by boosting the three momentum at rest, will be
characterized by the mass-shell condition
\begin{gather*}
\det {\mathbf p}=\big(p^0\big)^2-|\underline{p}|^2=m^2 .
\end{gather*}
We thus have that in three-dimensional Minkowski space the extended momentum space of a~relativistic point particle can
be identif\/ied with $\mathfrak{sl}(2,\mathbb{R})$ and the physical momenta belong to orbits of ${\rm SL}(2,\mathbb{R})$
on $\mathfrak{sl}(2,\mathbb{R})$ which determine the mass-shell.

\looseness=-1
 We can now have an intuitive characterization of the momentum space of a~moving conical defect in three
dimensions (for more formal discussions we refer the reader to~\cite{Arzano, MATS}).
As discussed above the momentum at rest of a~conical defect can be parametrized by a~rotation $\overline{{\mathbf
g}}=e^{4 \pi G m\gamma_0}\in {\rm SL}(2,\mathbb{R})$, i.e.~a group element rather than the vector $\overline{{\mathbf
p}} = m\gamma_0$.
The action of a~Lorentz boost on the momentum at rest will be described just by an action of ${\rm SL}(2,\mathbb{R})$ on
itself
\begin{gather}
{\bf g} ={\bf h}^{-1} \overline{{\mathbf g}} {\bf h}
\label{eq:conjug}
\end{gather}
and thus physical momenta of a~defect of mass~$m$ will be given by elements of ${\rm SL}(2,\mathbb{R})$ belonging to the
{\it conjugacy class} of a~rotation by an angle $4 \pi G m$.
Therefore, when gravity is switched on, the extended momentum space of a~point particle is given by the group manifold
${\rm SL}(2,\mathbb{R})$ (to be contrasted with the vector space $\mathfrak{sl}(2,\mathbb{R})$ in the ordinary Minkowski
case) while its physical momentum space is given by the action by conjugation of ${\rm SL}(2,\mathbb{R})$ {\it on the
momentum at rest} $\overline{{\mathbf g}}=e^{4 \pi G m\gamma_0}$ (to be contrasted with the adjoint action in the
ordinary case)~\cite{Arzano, MATS}.
In Section~\ref{section4.2} we will discuss how the transition from vector valued momenta to group valued momenta viz.
from equation~\eqref{eq:adjoint} to~\eqref{eq:conjug} can be understood from a~mathematical point of view in terms of
a~{\it lift} morphism between functions on Minkowski space to functions on ${\rm SL}(2,\mathbb{R})$.
Physically speaking the transition from~\eqref{eq:adjoint} to~\eqref{eq:conjug} simply ref\/lects the fact that the phase
space of a~point particle coupled to $2+1$ gravity is the cotangent bundle of ${\rm SL}(2,\mathbb{R})$ with the latter
describing the momentum degrees of freedom\footnote{Very similar structures arise in loop quantum gravity and group
f\/ield theory where the conf\/iguration space is given by a~Lie group, see e.g.~\cite{Baratin:2010nn,Baratin:2010wi, Noui}.}
of the particle~\cite{MATS}.
In the next section we give a~detailed account of the group momentum space of the particle and of the classif\/ication of
the conjugacy classes describing ``on-shell'' momenta.

\subsection{Conjugacy classes as deformed mass-shell}\label{section2.2}

In order to introduce the conjugacy classes of the group ${\rm SL}(2,\mathbb{R})$ we start by recalling the
expansion of the generic group element ${\mathbf g} \in {\rm SL}(2,\mathbb{R})$ in terms of the unit and the~$\gamma$
matrices:
\begin{gather}
\label{cartesian}
{\mathbf g}=u\mathbbm{1}+\xi^{\mu}\gamma_{\mu} ,
\end{gather}
which can be written in the following matrix form:
\begin{gather}
{\mathbf g}=
\begin{bmatrix}
u+\xi^2  & \xi^1+\xi^0
\\
\xi^1-\xi^0 &  u-\xi^2
\end{bmatrix}.
\label{eq:momematr}
\end{gather}
The unit determinant condition on this matrix gives $u^2-(\xi^1)^2-(\xi^2)^2+(\xi^0)^2=1$ and thus the~$u$ parameter can
be expressed in terms of the other free parameters and can be either positive or negative.
If we were considering instead of ${\rm SL}(2,\mathbb{R})$ the proper orthocronous Lorentz group ${\rm SO}_+(1,2)$ positive
and negative values of~$u$ would be identif\/ied.
This ref\/lects the fact that ${\rm SL}(2,\mathbb{R})$ is a~{\it double cover} of ${\rm SO}_+(1,2)$.
Now the action by conjugation of ${\rm SL}(2,\mathbb{R})$ on itself leaves the trace invariant, therefore the number~$u$
is an invariant under action by conjugation.
In particular, the group ${\rm SL}(2,\mathbb{R})$ is composed by f\/ive dif\/ferent subclasses, called {\it conjugacy
classes}, which are invariant under conjugation~\cite{Baez}.
These subclasses can be described by the eigenvalues of the matrix~\eqref{eq:momematr}.
The secular equation for the generic element of ${\rm SL}(2,\mathbb{R})$ reads
\begin{gather*}
\det [{\mathbf g}-\lambda\mathbbm{1} ]=
\begin{vmatrix}
u+\xi^2-\lambda  & \xi^1+\xi^0
\\
\xi^1-\xi^0 &  u-\xi^2-\lambda
\end{vmatrix}
=0  ,
\end{gather*}
whose solutions are given~by
\begin{gather*}
\lambda= \frac{\text{Tr}({\mathbf g}) \pm \sqrt{[\text{Tr}({\mathbf g})]^2-4}}{2}
.
\end{gather*}
We can now classify group elements according to their trace, in fact we see that the above equation has dif\/ferent
behaviours according to the value of the discriminant $[\text{Tr}({\mathbf g})]^2-4$, in particular the sign of this
term will determine the dif\/ferent sets of solutions, i.e.~group elements:
\begin{itemize}\itemsep=0pt
\item If $|\text{Tr}({\mathbf g})|<2$ ($|u|<1$), then ${\mathbf g}$ is called {\it elliptic} (the geometry of the
parameter space is given by the elliptic hyperboloid in Fig.~\ref{tsheets} in Appendix~\ref{appendixA}) and it is conjugate to {\it rotations}    ${\mathbf r} \in {\rm SL}(2,\mathbb{R}$):
\begin{gather*}
\mathbf{r}=
\begin{bmatrix}
\cos(\alpha) &  \sin(\alpha)
\\
-\sin(\alpha) &  \cos(\alpha)
\end{bmatrix},
\end{gather*}
where $\alpha \in [0,2\pi)$ and $-2<\text{Tr}({\mathbf r})< 2$.

\item If $|\text{Tr}({\mathbf g})|>2$ ($|u|>1$), then ${\mathbf g}$ is called {\it hyperbolic} (the geometry of the
parameter space is given by the hyperbolic hyperboloid in Fig.~\ref{des} in Appendix~\ref{appendixA}) and it is conjugate to {\it boosts}
${\mathbf b}$ or {\it antiboosts} $\tilde{{\mathbf b}}$:
\begin{gather*}
{\mathbf b}=
\begin{bmatrix}
e^{\beta} &  0
\\
0  & e^ {-\beta}
\end{bmatrix}
,
\qquad
\tilde{{\mathbf b}}=
\begin{bmatrix}
- e^{\beta} &  0
\\
0 &  -e^ {-\beta}
\end{bmatrix},
\end{gather*}
where $\beta \in [0,\infty)$, $\text{Tr}({\mathbf b}) > 2$ and $\text{Tr}(\tilde{{\mathbf b}}) < -2$.
\item If $|\text{Tr}({\mathbf g})|=2$ ($|u|=1$), then ${\mathbf g}$ is called {\it parabolic} (the geometry of the
parameter space is given by the cone in Fig.~\ref{conem} in Appendix~\ref{appendixA}) and it is conjugate to the {\it shears}~${\mathbf s}$ or
{\it antishears}~$\tilde{{\mathbf s}}$:
\begin{gather*}
{\mathbf s}=
\begin{bmatrix}
1  & \gamma
\\
0 &  1
\end{bmatrix},
\qquad
\tilde{{\mathbf s}}=
\begin{bmatrix}
-1  & \gamma
\\
0 &  -1
\end{bmatrix},
\end{gather*}
where $\gamma=0,\pm 1$, $\text{Tr}({\mathbf s}) = 2$ and $\text{Tr}(\tilde{{\mathbf s}}) = -2$.
\end{itemize}

To make contact with the ordinary classif\/ication of particles according to the sign of their mass squared, we have here
that a~group element obtained by conjugating the momentum at rest of a~gravitating particle
\begin{gather*}
\overline{{\mathbf g}}=\cos (4 \pi G m)\mathbbm{1}+\sin (4 \pi G m)\gamma_0=
\begin{bmatrix}
\cos (4 \pi G m) &  \sin (4 \pi G m)
\\
-\sin (4 \pi G m) &  \cos (4 \pi G m)
\end{bmatrix},
\end{gather*}
belongs to the class of elliptic transformations, which represent massive particles.
Analogously, hyperbolic elements will represent tachyons and the parabolic elements photons as we can deduce by taking
the $G\rightarrow 0$ limit and checking that for the former we have a~negative mass squared, while for the latter the
mass is vanishing.

The deformed mass-shell condition assigning a~group element to the conjugacy class describing massive particles can be
obtained by putting a~constraint on its trace which, in light of what we said above, will be given by
\begin{gather}
\label{conjcond}
\frac{1}{2}\text{Tr}({\mathbf g})=\cos(4 \pi G m),
\end{gather}
with $0<m\le\frac{1}{8G}=\frac{\kappa\pi}{2}$ where we def\/ined a~new deformation parameter with dimensions of mass
$\kappa=\frac{1}{4 \pi G}$ for later convenience.
This range of masses implies a~choice of positive sign of the variable~$u$ discussed above.
This choice of sign, widely adopted in the literature, is mathematically equivalent to considering the quotient
${\rm SO}_+(1,2)={\rm SL}(2,\mathbb{R}) / Z_2$, where $Z_2$ is the cyclic group.

In what follows, we will give concrete examples of group parametrizations and their (deformed) mass-shell, before moving
on to the description of how relativistic symmetries are implemented for such particles.

\section[Coordinates and symmetries on the ${\rm SL}(2,\mathbb{R})$ momentum space]{Coordinates and symmetries on the $\boldsymbol{{\rm SL}(2{,}\mathbb{R})}$ momentum space}\label{section3}

In this section we recall some known parametrizations of ${\rm SL}(2,\mathbb{R})$ group elements/momenta and introduce
others which will help us get insight on the kinematical properties associated to each set of coordinates.
Furthermore, we will implement Lorentz transformations in the picture discussing how these are realized for each choice
of parameterization adopted.

\subsection{Cartesian coordinates}\label{section3.1}

{\it Cartesian coordinates} on the ${\rm SL}(2,\mathbb{R})$ group manifold are the most used in the
literature.
They are based on the parametrization~\eqref{cartesian} which we introduced in Section~\ref{section2.2}
\begin{gather*}
{\bf g}=u\mathbbm{1}+\xi^{\mu}\gamma_{\mu}  .
\end{gather*}
We def\/ine
\begin{gather*}
 u=\sqrt{1+|\vec{p}_{\kappa}|^2}, \qquad
\xi^0=p_{\kappa}^0, \qquad
\xi^1= p^1_{\kappa}, \qquad
\xi^2= p^2_{\kappa},
\end{gather*}
where we introduced the notation $p_{\kappa}^{\mu} \doteq \frac{p^{\mu}}{\kappa}$, $\mu=0, 1, 2$, so that the group
parameters $\xi^{\mu}$ are dimensionless and the coordinates $p^{\mu}$ have the dimensions of energy.
The deformed mass-shell constraint~\eqref{conjcond} in Cartesian coordinates reads
\begin{gather*}
\frac{1}{2}\text{Tr}({\mathbf g})= u =\sqrt{1+|\vec{p}_{\kappa}|^2}=\cos m_{\kappa}  ,
\end{gather*}
which after easy manipulations gives
\begin{gather}
|\vec{p}_{\kappa}|^2=-\sin^2m_{\kappa} \  \Rightarrow \  \big(p^0\big)^2=|\underline{p}|^2+k^2\sin^2m_{\kappa}  .
\label{masss}
\end{gather}
In the limit $\kappa \to \infty$ the dispersion relation~\eqref{masss} reproduces the usual mass-shell relation valid
for an ordinary (f\/lat) momentum space of a~massive particle, i.e.~$(p^0)^2=|\underline{p}|^2+m^2$.
The deformation given by the gravity ef\/fects, using this parametrization, appears as a~renormalization of the mass which
is no longer~$m$ but $\mathcal{M} \doteq \kappa \sin m_{\kappa}$.

\subsubsection{Boosts for Cartesian coordinates}\label{section3.1.1}

We show now how momenta described by cartesian coordinates transform under Lorentz boosts.
Without loss of generality we consider a~Lorentz boost in $\gamma_1$-direction.
This is represented by the group element ${\mathbf b} = e^{\frac{1}{2}\eta\gamma_2}$ where~$\eta$ is the boost rapidity.
The action of ${\rm SL}(2, \mathbb{R})$ on itself is realized by conjugation
\begin{gather*}
{\mathbf g}'={\mathbf b}\triangleright {\bf g}={\bf b}^{-1}{\bf g}{\bf b},
\end{gather*}
which in components will read
\begin{gather*}
\sqrt{1+|\vec{p}_{\kappa}^{\,\prime}|^2}\mathbbm{1}+p'^{\mu}_{\kappa}\gamma_{\mu}=
e^{-\frac{1}{2}\eta\gamma_2}\bigl[\sqrt{1+|\vec{p}_{\kappa}|^2}\mathbbm{1}+p^{\mu}_{\kappa}\gamma_{\mu}\bigl]
e^{\frac{1}{2}\eta\gamma_2}  .
\end{gather*}
Writing down the product on the right hand side above, using the multiplication properties of the~$\gamma$ matrices and
putting together the terms proportional to the same~$\gamma$ matrix we obtain the expressions for the boosted parameters
$p'^{\mu}$ in terms of the old parameters $p^{\mu}$
\begin{gather*}
 |\vec{p}\,{}'|^2=|\vec{p}\,|^2,\qquad
p'^0= p^0 \cosh\eta- p^1 \sinh\eta, \qquad
p'^1=p^1\cosh\eta-p^0\sinh\eta,  \qquad
p'^2= p^2.
\end{gather*}
We thus obtain the {\it ordinary} action of a~Lorentz boost in the $x^1$ direction.

\subsection{Exponential coordinates}\label{section3.2}

 This set of coordinates is widely used in the works on Euclidean models (see e.g.~\cite{oriti}) in which
${\rm SU}(2)$ is considered instead of ${\rm SL}(2,\mathbb{R})$.
The parametrization is def\/ined as follows (see also~\cite{Sasai:2009jm})
\begin{gather*}
 u=\cosh|\vec{k}_{\kappa}|,
\qquad
\xi^0= \frac{\sinh|\vec{k}_{\kappa}|}{|\vec{k}_{\kappa}|}k^0_{\kappa},
\qquad
\xi^1= \frac{\sinh|\vec{k}_{\kappa}|}{|\vec{k}_{\kappa}|}k^1_{\kappa},
\qquad
\xi^2= \frac{\sinh|\vec{k}_{\kappa}|}{|\vec{k}_{\kappa}|}k^2_{\kappa},
\end{gather*}
It is trivial to check that for this parametrization the determinant constraint is satisf\/ied
\begin{gather*}
\det {\mathbf g}=
\cosh^2|\vec{k}_{\kappa}|-\frac{\sinh^2|\vec{k}_{\kappa}|}{|\vec{k}_{\kappa}|^2}|\vec{k}_{\kappa}|^2=\cosh^2|\vec{k}_{\kappa}|-\sinh^2|\vec{k}_{\kappa}|=1,
\end{gather*}
and a~generic group element can be written in terms of $\vec{k}$-coordinates as
\begin{gather*}
{\bf g}=\cosh|\vec{k}_{\kappa}|\mathbbm{1}+
\frac{\sinh|\vec{k}_{\kappa}|}{|\vec{k}_{\kappa}|}k^{\mu}_{\kappa}\gamma_{\mu}=e^{k^{\mu}_{\kappa}\gamma_{\mu}} ,
\end{gather*}
where the last equality shows that this parametrization is equivalent to considering the exponential of the
$\frak{sl}(2,\mathbb{R})$ generators, from this the name ``exponential coordinates''.
The conjugacy class constraint this time leads to the mass-shell condition
\begin{gather*}
\cosh|\vec{k}_{\kappa}|=\cos m_{\kappa} \ \Leftrightarrow  \ \cosh|\vec{k}_{\kappa}|=\cosh i m_{\kappa} \ \Rightarrow \
\big(k^0\big)^2=E^2_{[\vec{k}]}=|\underline{k}|^2+m^2,
\end{gather*}
which is just an ordinary mass-shell relation where, however, the values of the mass~$m$ are limited~by
$\kappa\frac{\pi}{2}$ due to the limits on the def\/icit angle.

\subsubsection{Boosts for exponential coordinates}\label{section3.2.1}

  As in the previous case we consider the action by conjugation of the Lorentz boost in a~$\gamma_1$-direction
\begin{gather*}
{\mathbf g}'={\mathbf b}\triangleright {\mathbf g} \ \Rightarrow \
e^{k'^{\mu}_{\kappa}\gamma_{\mu}}=e^{-\frac{1}{2}\eta\gamma_2}e^{k^{\mu}_{\kappa}\gamma_{\mu}}e^{\frac{1}{2}\eta\gamma_2}.
\end{gather*}
In analogy with the previous case we can f\/ind an explicit expression for the boosted parameters using the matrix
expansion of the exponentials
\begin{gather*}
 |\vec{k}\,{}'|^2=|\vec{k}|^2,
\qquad
k'^0=k^0\cosh\eta-k^1\sinh\eta,
\qquad
k'^1=k^1\cosh \eta - k^0\sinh \eta,
\qquad
k'^2=k^2
\end{gather*}
and thus also the exponential coordinates transform as ordinary momenta in Minkowski space under a~boost.

\subsection{Time-symmetric parametrization: Euler coordinates}\label{section3.3}

\begin{figure}[htbp] \centering
\includegraphics[width=7cm]{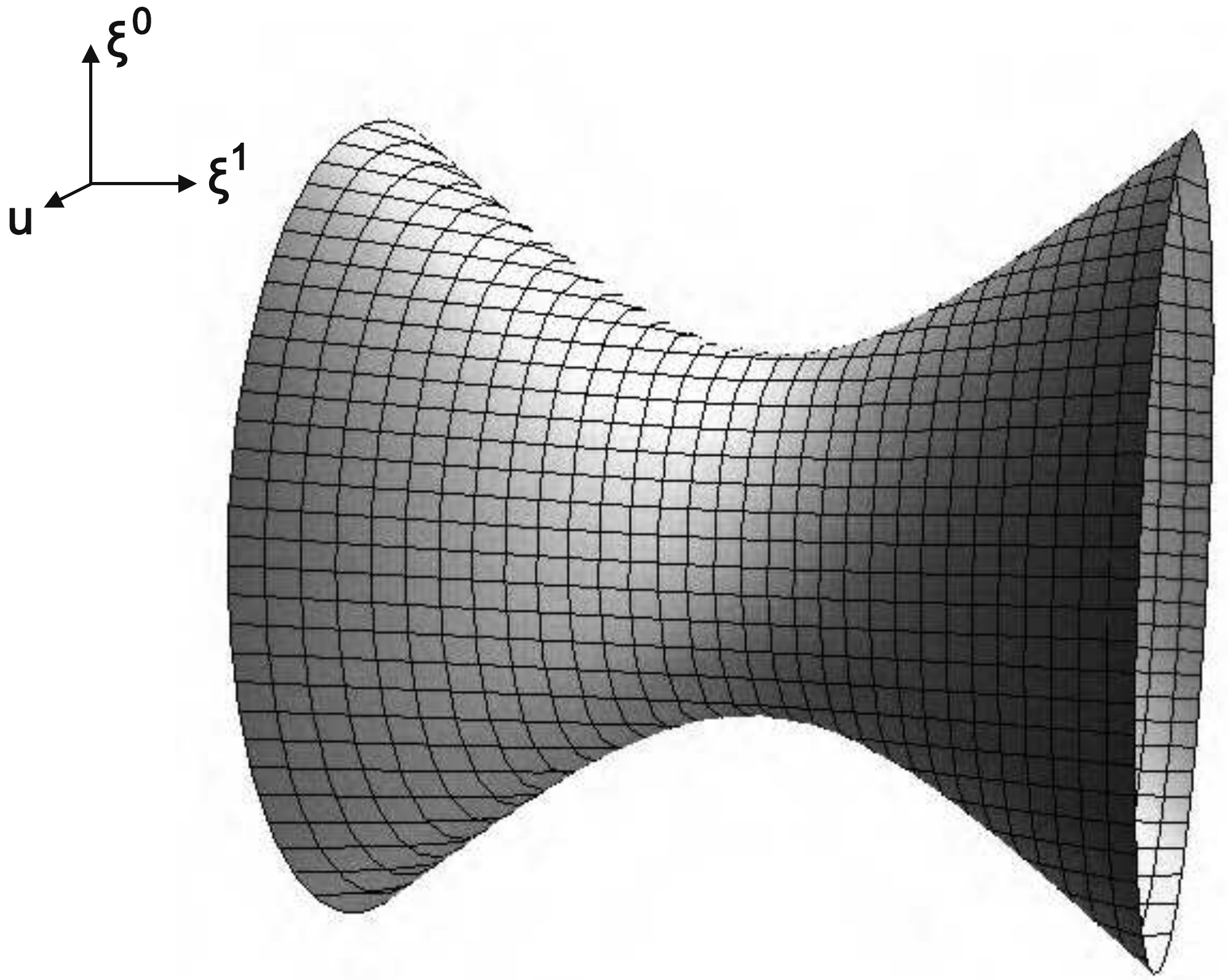}
\caption{Group manifold ${\rm
SL}(2,\mathbb{R})$ embedded in $\mathbb{R}^ {2,2}$ with $\xi^2$ coordinate suppressed.
The grid lines on the AdS are the Euler angles~$\rho$ and~$\chi$, the third angle~$\phi$ does not appear in the picture
but we have to imagine it as a~rotation in the suppressed dimension~$\xi^2$.}\label{euldes}
\end{figure}

  In this section we develop a~set of coordinates on ${\rm SL}(2,\mathbb{R})$ based on a~choice of energy and
linear momentum introduced in earlier works on the coupling of point particles to $2+1$
gravity~\cite{MATS,'tHooft:1996uc,Welling:1997qz}.
The idea is to describe the momenta of a~gravitating particle using a~parametrization in terms of angles (closely
related to the Euler angles parametrizing ${\rm SU}(2)$)
\begin{gather*}
 u=\cosh\chi\cos\rho,\qquad
\xi^0=\cosh\chi\sin\rho,\qquad
\xi^1=\sinh\chi\cos\phi,\qquad
\xi^2=\sinh\chi\sin\phi ,
\end{gather*}
where $\chi \in [0,\infty)$ and~$\rho,\phi \in [0,2\pi)$ (see Fig.~\ref{euldes}).

Now, using this set of coordinates, we can rewrite the expansion of the group element ${\mathbf g}$ in terms
of the~$\gamma$ matrices as~\cite{MATS}:
\begin{gather*}
{\bf g}= e^{\frac{1}{2}(\rho + \phi)\gamma_0}e^{\chi \gamma_1}e^{\frac{1}{2}(\rho - \phi)\gamma_0}.
\end{gather*}
This choice of coordinates shows clearly that the Lorentz transformation represented by ${\mathbf g}$ is decomposed in
a~spatial rotation of angle $\rho + \phi$ in the plane $\gamma_1-\gamma_2$, a~boost in $\gamma_2$ direction with
the~$\chi$ parameter and another rotation in the same plane of the previous but by angle $\rho - \phi$.

\begin{figure}[htbp]\centering
\includegraphics[width=10cm]{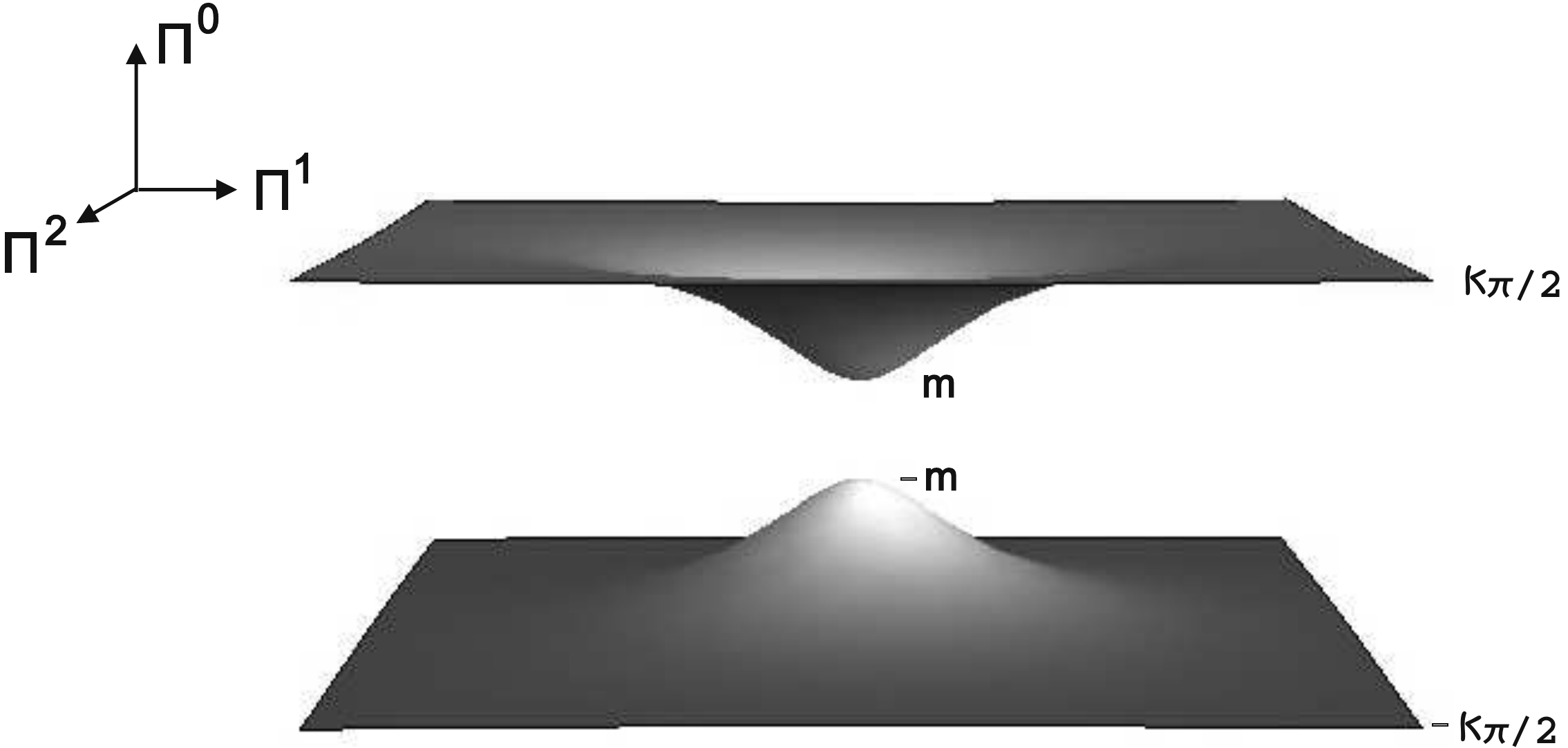}

\caption{Deformed mass-shell in terms of the
parameters $\Pi^{\mu}$, $\mu=0,1,2$.
Note that when~$m$ reaches its maximum value $\frac{\kappa \pi} {2}$ the momentum space collapses into two parallel
inf\/inite planes at $\Pi^0=\pm \frac{\kappa \pi}{2}$.}\label{momespace}
\end{figure}

In order to interpret our coordinates as momenta we rewrite the expansion of the group element using~$\rho$
and~$\chi$ as dimensionful parameters $\rho \rightarrow \rho_{\kappa}$ and $\chi \rightarrow \chi_{\kappa}$, obtaining
\begin{gather*}
u=\cosh\chi_{\kappa}\cos\rho_{\kappa}, \qquad
\xi^0=\cosh\chi_{\kappa}\sin\rho_{\kappa}, \qquad
\xi^1=\sinh\chi_{\kappa}\cos\phi, \qquad
\xi^2=\sinh\chi_{\kappa}\sin\phi.
\end{gather*}
Now, as in the previous cases, if we impose the condition which projects the group element into the conjugacy class of
rotations, i.e.~$\frac{1}{2}\text{Tr}({\mathbf g})=\cos m_{\kappa}$, we obtain
\begin{gather}
\cosh\chi_{\kappa}\cos\rho_{\kappa}=\cos m_{\kappa},
\label{eq:masshell2}
\end{gather}
which is the deformed mass-shell relation in our new coordinates.
If we take the limit $\kappa \rightarrow \infty$
\begin{gather*}
\left[1+\frac{\chi_{\kappa}^2}{2}+\mathcal{O}\big(\kappa^{-4}\big)\right]\left[1-\frac{\rho_{\kappa}^2}{2}+\mathcal{O}\big(\kappa^{-4}\big)\right]
=1-\frac{m_{\kappa}^2}{2}+\mathcal{O}\big(\kappa^{-4}\big),
\end{gather*}
which, simplifying the leading terms and identifying the terms proportional to $\kappa^{-2}$, gives
\begin{gather*}
\rho^2=\chi^2+m^2  .
\end{gather*}
Such equation reproduces an ordinary relativistic mass-shell relation if one identif\/ies~$\rho$ and~$\chi$ respectively
with the energy and the norm of the spatial momentum.
This leads us to def\/ine the following momenta for the gravitating particles
\begin{gather*}
 \Pi^0 \doteq \rho, \qquad
\Pi^1 \doteq \chi\cos\phi=|\underline{\Pi}|\cos\phi, \qquad
\Pi^2 \doteq \chi\sin\phi=|\underline{\Pi}|\sin\phi.
\end{gather*}
In terms of the new $\vec{\Pi}$-coordinates we have
\begin{gather*}
 u=\cosh|\underline{\Pi}_{\kappa}|\cos\Pi_{\kappa}^0, \!\!\qquad
\xi^0= \cosh|\underline{\Pi}_{\kappa}|\sin\Pi_{\kappa}^0, \!\!\qquad
\xi^1= \frac{\sinh|\underline{\Pi}_{\kappa}|}{|\underline{\Pi}_{\kappa}|}\Pi_{\kappa}^1, \!\!\qquad
\xi^2= \frac{\sinh|\underline{\Pi}_{\kappa}|}{|\underline{\Pi}_{\kappa}|}\Pi_{\kappa}^2.\!
\end{gather*}
Using this parametrization, the generic group element can be written as
\begin{gather}
{\mathbf g}=
e^{\frac{\Pi^0_{\kappa}}{2}\gamma_0} e^{\Pi^1_{\kappa}\gamma_1+\Pi^2_{\kappa}\gamma_2} e^{\frac{\Pi^0_{\kappa}}{2}\gamma_0},
\label{eq:piform}
\end{gather}
as shown in detail in Appendix~\ref{appendixB}. The mass-shell relation~\eqref{eq:masshell2} can be rewritten in terms of the~$\vec{\Pi}$ momenta as
\begin{gather*}
\cosh|\underline{\Pi}_{\kappa}|\cos\Pi_{\kappa}^0=\cos m_{\kappa},
\end{gather*}
which solved for the energy gives
\begin{gather*}
\Pi^0=E_{[\vec{\Pi}]}=\pm\kappa\arccos \left(\frac{\cos m_{\kappa}}{\cosh |\underline{\Pi}_{\kappa}|}\right),
\end{gather*}
showing that the energy $\Pi^0$ is limited by the range values $\Pi^0 \in [m,\frac{\kappa\pi}{2}]$ (see Fig.~\ref{momespace}).
This was somewhat an expected result, since the energy in this set of coordinates is proportional to the def\/icit angle
but we will have more to say about this feature when considering Lorentz transformations in the next sections.

\subsubsection{Boosts on Euler coordinates}

Again, we can compute the expression ${\mathbf g}'={\mathbf b}\triangleright {\bf g}$ using the same steps of the
previous cases.

\begin{figure}[htbp] \centering

\includegraphics[width=10cm]{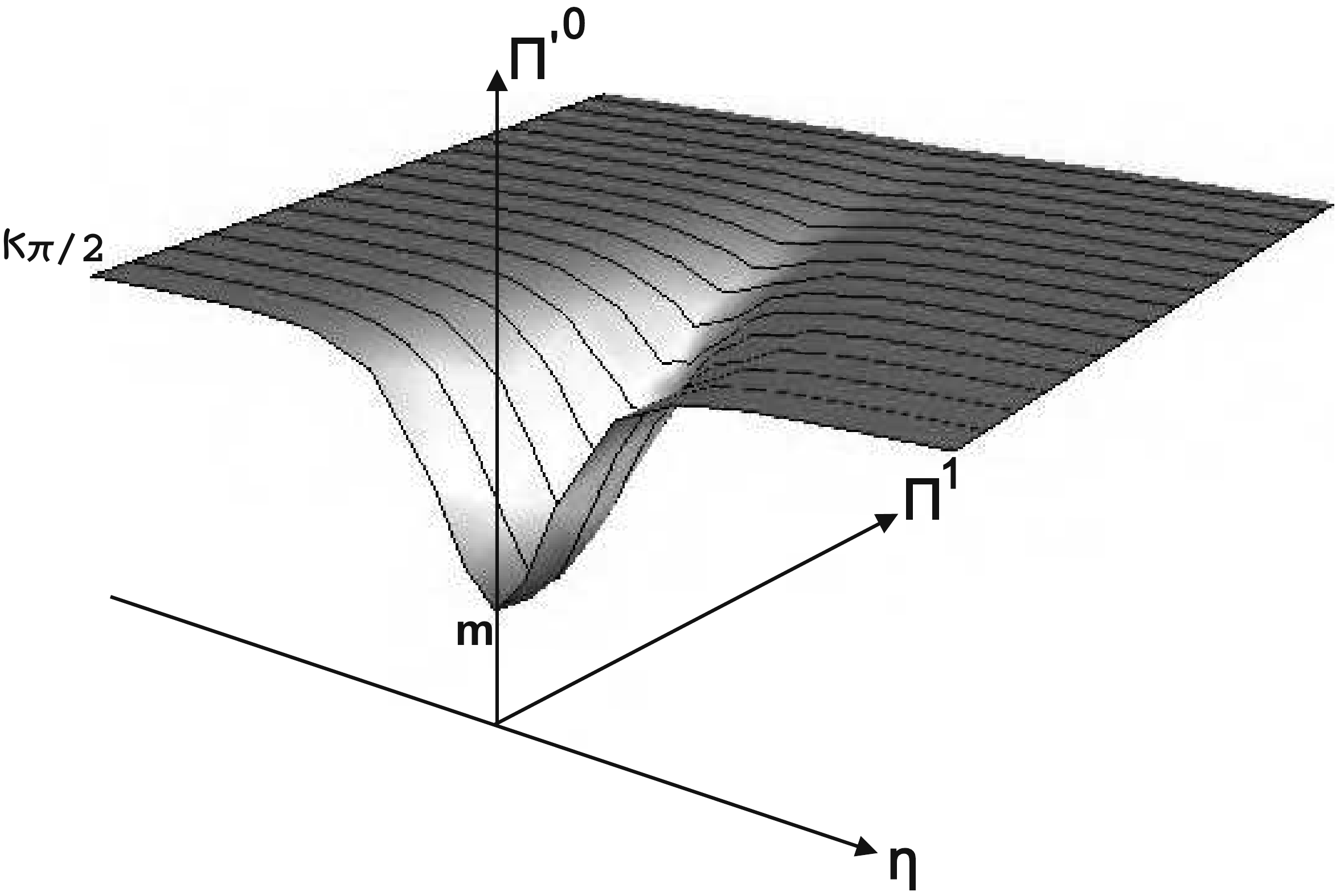}

\caption{Boosted energy $\Pi'^0$ calculated for $\Pi^2=0$.
The value of the mass is $m=\frac{\kappa\pi}{4}$.
For $m=\frac{\kappa\pi}{2}$ this surface becomes an inf\/inite plane at $\Pi'^0=\frac{\kappa \pi}{2}$.}\label{boostsh}
\end{figure}

This leads us to write the following system
\begin{gather*}
\cosh|\underline{\Pi}'_{\kappa}| \cos\Pi'^0_{\kappa}=\cosh|\underline{\Pi}_{\kappa}| \cos\Pi^0_{\kappa},\\
\cosh|\underline{\Pi}'_{\kappa}|\sin\Pi'^0_{\kappa}=\cosh|\underline{\Pi}_{\kappa}|\sin\Pi^0_{\kappa} \cosh\eta-
\frac{\sinh|\underline{\Pi}_{\kappa}|} {|\underline{\Pi}_{\kappa}|}\Pi^1_{\kappa} \sinh\eta,\\
\frac{\sinh|\underline{\Pi}'_{\kappa}|}{|\underline{\Pi'}_{\kappa}|}\Pi'^1_{\kappa}=
\frac{\sinh|\underline{\Pi}_{\kappa}|}{|\underline{\Pi}_{\kappa}|}\Pi^1_{\kappa} \cosh\eta-
\cosh|\underline{\Pi}_{\kappa}|\sin\Pi^0_{\kappa}\sinh\eta,\\
\frac{\sinh|\underline{\Pi}'_{\kappa}|} {|\underline{\Pi'}_{\kappa}|}\Pi'^2_{\kappa}=
\frac{\sinh|\underline{\Pi}_{\kappa}|}{|\underline{\Pi}_{\kappa}|}\Pi^2_{\kappa},
\end{gather*}
whose solution is given by
\begin{gather}
 \Pi'^0=\kappa \arctan \biggl(\tan \Pi^0_{\kappa} \cosh \eta -\frac{\tanh
|\underline{\Pi}_{\kappa}|}{|\underline{\Pi}|} \Pi^1\sinh \eta\biggl),\nonumber\\
\Pi'^1=\Pi^1 \cosh \eta - \frac{|\underline{\Pi}|}{\tanh |\underline{\Pi}_{\kappa}|}\sin \Pi^0_{\kappa} \sinh \eta,\label{eq:boob}\\
\Pi'^2=\Pi^2.\nonumber
\end{gather}
We thus have that Euler coordinates on the ${\rm SL}(2,\mathbb{R})$ provide a~{\it non-linear realization} of Lorentz
boosts (see Fig.~\ref{boostsh}).
Obviously in the limit $\kappa\rightarrow\infty$ the transformations~\eqref{eq:boob} become just ordinary Lorentz
boosts.
Consider now a~particle initially at rest (namely $\Pi^0=m$, $\Pi^1=\Pi^2=0$), the system of equations~\eqref{eq:boob}
simplif\/ies to
\begin{gather*}
\Pi'^0=\kappa \arctan \bigl(\tan m_{\kappa} \cosh \eta \bigl), \qquad
\Pi'^1=- m \sinh \eta, \qquad \Pi'^2= 0.
\end{gather*}

We see that in the limit when the boost parameter goes to inf\/inity, $\eta \to \infty$, the energy does not
grow arbitrarily but saturates at the value~$\frac{\kappa\pi} {2}$.
We thus have that the non-linear realization of Lorentz boosts given in Euler coordinates is just an example of deformed
or {\it doubly special relativity}~\cite{AmelinoCamelia:2000mn,AmelinoCamelia:2010pd} in which we have a~maximum energy
built in the deformed structure of Lorentz transformations.

\section[The quantum double of ${\rm SL}(2,\mathbb{R})$]{The quantum double of $\boldsymbol{{\rm SL}(2,\mathbb{R})}$} \label{section4}

In this section we show how the transition from a~Minkowski to ${\rm SL}(2, \mathbb{R})$ momentum
space translates for the structure of relativistic symmetries in a~deformation of the Poincar\'e group to the quantum
double of ${\rm SL}(2, \mathbb{R})$, a~quantum group denoted as $\mathcal{D}({\rm SL}(2, \mathbb{R}))$.
We start by recalling the def\/inition and Hopf algebra properties of $\mathcal{D}({\rm SL}(2, \mathbb{R}))$.
For a~gentle introduction to the necessary notions of Hopf algebras we refer the reader to~\cite{fuchs} while a~rigorous
introduction to $\mathcal{D}({\rm SL}(2, \mathbb{R}))$ can be found in~\cite{Koornwinder:1996uq}.

The quantum double $\mathcal{D}({\rm SL}(2, \mathbb{R}))$ is def\/ined, as a~vector space, by the tensor
product~\cite{joung}
\begin{gather*}
\mathcal{D}({\rm SL}(2, \mathbb{R}))=C({\rm SL}(2, \mathbb{R})) \otimes \mathbbm{C}({\rm SL}(2, \mathbb{R})),
\end{gather*}
where $C({\rm SL}(2, \mathbb{R}))$ is the space of complex functions on ${\rm SL}(2, \mathbb{R})$ and $\mathbbm{C}({\rm SL}(2, \mathbb{R}))$ the group algebra of ${\rm SL}(2, \mathbb{R})$ which, roughly speaking, can be thought of as
a~vector space comprised by the elements of the group itself.
Let us denote an element of the double as $(f \otimes \mathbf{g}) \in \mathcal{D}({\rm SL}(2, \mathbb{R}))$, the Hopf
algebra structure is given~by
\begin{alignat*}{3}
 & \text{product:} \quad &&
(f_1 \otimes \mathbf{g})\cdot (f_2 \otimes \mathbf{h})=(f_1 \cdot \text{ad}_{\mathbf{g}}f_2 \otimes \mathbf{gh}), & \\
& \text{co-product:}  \qquad &&
\Delta_{\mathcal{D}({\rm SL}(2,\mathbb{R}))}(f \otimes \mathbf{g}) = \sum\limits_{(f)}(f_{(1)}\otimes \mathbf{g})
\otimes (f_{(2)}\otimes \mathbf{g}), &\\
& \text{unit:} \qquad &&  (1 \otimes e), &\\
& \text{co-unit:}  \qquad &&  \epsilon (f \otimes \mathbf{g})=f(e), &\\
& \text{antipode:}  \qquad && S(f \otimes \mathbf{g})=\big(\iota\text{ad}_{\mathbf{g}^{-1}}f \otimes \mathbf{g}^{-1}\big) , &\\
& \text{complex conjugate:} \qquad &&
(f \otimes \mathbf{g})^*=\big(\overline{\text{ad}_{\mathbf{g}^{-1}}f} \otimes \mathbf{g}^{-1}\big),  &
\end{alignat*}
where $\text{ad}_{\mathbf{g}}f(\mathbf{h}) \overset{\mathrm{def}}{=}
f(\mathbf{g}\mathbf{h}\mathbf{g}^{-1})$ and $\iota f(\mathbf{g}) \overset{\mathrm{def}}{=}
f(\mathbf{g^{-1}})$.
The explicit expression for the co-product on $C({\rm SL}(2, \mathbb{R}))$ is given~by
\begin{gather*}
\Delta_{\mathcal{C}({\rm SL}(2, \mathbb{R}))} f(\mathbf{g} \otimes \mathbf{h}) =
\sum\limits_{(f)}f_{(1)}(\mathbf{g})f_{(2)}(\mathbf{h})=f(\mathbf{g}\mathbf{h}) ,
\end{gather*}
telling us that for such space the co-product is simply a~way of extending the notion of function on the group to
a~tensor product of the latter in a~way compatible with the composition of the group.

\subsection[$\mathcal{D}({\rm SL}(2, \mathbb{R}))$ as a deformation of $\mathbbm{C}({\rm ISL}(2, \mathbb{R}))$]{$\boldsymbol{\mathcal{D}({\rm SL}(2, \mathbb{R}))}$ as a deformation of $\boldsymbol{\mathbbm{C}({\rm ISL}(2, \mathbb{R}))}$}

Here we show how the quantum double of ${\rm SL}(2, \mathbb{R})$ can be interpreted as a~{\it deformation} of
the group algebra of the three dimensional Poincar\'e group, or inhomogenous Lorentz group ${\rm ISL}(2,\mathbb{R})$.
The group algebra $\mathbbm{C}({\rm ISL}(2,\mathbb{R}))$ is def\/ined as a~vector space by the tensor product
\begin{gather*}
\mathbbm{C}({\rm ISL}(2, \mathbb{R}))=\mathbbm{C}\big(\mathbb{R}^{2,1} \rtimes {\rm SL}(2, \mathbb{R})\big) =
\mathbbm{C}\big(\mathbb{R}^{2,1}\big) \otimes \mathbbm{C}({\rm SL}(2, \mathbb{R})),
\end{gather*}
where $\mathbb{R}^{2,1}$ is the group of translations which we denote as $\mathcal{T}_x$ with $x \in
\mathbb{R}^{2,1}$.
The group algebra $\mathbbm{C}(\mathbb{R}^{2,1})$ is the set of elements $\int d^3x \tilde{f}(x)\mathcal{T}_x$, with
$\tilde{f}(x)$ a~function (or distribution) with compact support.
$\mathbbm{C}(\mathbb{R}^{2,1})$ can be identif\/ied with a~subalgebra $C(\mathbb{R}^{2,1})$ of functions on
$\mathbb{R}^{2,1}$.
This means that for any element $\int d^3x \tilde{f}(x)\mathcal{T}_x$ of the group algebra we can associate a~function
$f(p)=\int d^3x \tilde{f}(x) e^{-i p x} \in C(\mathbb{R}^{2,1})$ whose Fourier transform has compact support.
This amounts to identify
\begin{gather*}
\mathbbm{C}\big(\mathbb{R}^{2,1}\big) \simeq C \big(\mathbb{R}^{2,1}\big)
\
\Longleftrightarrow
\
\mathcal{T}_x \in \mathbbm{C}\big(\mathbb{R}^{2,1}\big)\to T_x(p)=e^{-i p x} \in C\big(\mathbb{R}^{2,1}\big) .
\end{gather*}
In conclusion, the group algebra $\mathbbm{C}({\rm ISL}(2, \mathbb{R}))$ can be identif\/ied with the tensor product
\begin{gather*}
\mathbbm{C}({\rm ISL}(2, \mathbb{R})) = \mathbbm{C}\big(\mathbb{R}^{2,1}\big) \otimes \mathbbm{C}({\rm SL}(2, \mathbb{R}))
\simeq C\big(\mathbb{R}^{2,1}\big) \otimes \mathbbm{C}({\rm SL}(2, \mathbb{R})) .
\end{gather*}
At this point it is evident that the dif\/ference with the quantum double is that we have $C(\mathbb{R}^{2,1})$ instead
of $C({\rm SL}(2, \mathbb{R}))$.
Thus the connection between $\mathbbm{C}({\rm ISL}(2, \mathbb{R}))$ and $\mathcal{D}({\rm SL}(2, \mathbb{R}))$ is
a~``lift'' or ``deformation map'' from $C(\mathbb{R}^{2,1})$ to $C({\rm SL}(2, \mathbb{R}))$.
This will turn out to be an algebra but not a~co-algebra morphism~\cite{joung} since the structure of the co-product for
$C(\mathbb{R}^{2,1})$ will be dif\/ferent than the one for $C({\rm SL}(2, \mathbb{R}))$.

To obtain this map we write down the Hopf algebra structure of $\mathbbm{C}({\rm ISL}(2, \mathbb{R}))$
\begin{alignat*}{3}
&  \text{product:}  \qquad &&
(f_1 \otimes \mathbf{g})\cdot (f_2 \otimes \mathbf{h})=(f_1 \cdot \text{R}_{\mathbf{g}}f_2 \otimes \mathbf{gh}), & \\
&  \text{co-product:}  \qquad &&
\Delta_{\mathbbm{C}({\rm ISL}(2, \mathbb{R}))}(f \otimes \mathbf{g}) = \sum\limits_{(f)}(f_{(1)}\otimes \mathbf{g})
\otimes (f_{(2)}\otimes \mathbf{g}), &\\
&  \text{unit:}  \qquad && (1 \otimes e), &\\
& \text{co-unit:} \qquad && \epsilon (f \otimes \mathbf{g})=f(e), &\\
& \text{antipode:}  \qquad &&
S(f \otimes \mathbf{g})=\big(\iota\text{R}_{\mathbf{g}^{-1}}f \otimes \mathbf{g}^{-1}\big), &\\
&  \text{complex conugate:}  \qquad &&
(f \otimes \mathbf{g})^*=\big(\overline{\text{R}_{\mathbf{g}^{-1}}f} \otimes \mathbf{g}^{-1}\big) ,&
\end{alignat*}
where the inverse map~$\iota$ is def\/ined as in the quantum double $\text{R}_{\mathbf{g}}f(\vec{p})
\overset{\mathrm{def}}{=} f(R(\mathbf{g}^{-1}) \vec{p})$ where $R(\mathbf{g})$ is a~vector representation
of ${\rm SL}(2, \mathbb{R})$.
The explicit expression for the co-product of $C(\mathbb{R}^{2,1})$ is given~by
\begin{gather}
\Delta_{\mathcal{C}(\mathbb{R}^{2,1})} f(\vec{p} \otimes \vec{q}) =
\sum\limits_{(f)}f_{(1)}(\vec{p})f_{(2)}(\vec{q})=f(\vec{p}+\vec{q})  ,
\end{gather}
which dif\/fers from the co-product for $C({\rm SL}(2, \mathbb{R}))$ since the composition law of the arguments of~$f$ is
abelian while in the co-product for $C({\rm SL}(2, \mathbb{R}))$ is not.
Notice however that the Hopf algebra structure of $\mathbbm{C}({\rm ISL}(2, \mathbb{R}))$ is very similar to the one of
the quantum double of ${\rm SL}(2, \mathbb{R})$.
In particular instead of the map $\text{ad}_{\mathbf{g}}(\cdot)$ we now have $\text{R}_{\mathbf{g}}(\cdot)$, thus to
pass from one Hopf algebra to the other we need a~connection between the two maps.
In other words we are looking for an algebra morphism~$\varphi$ such that
\begin{gather*}
 C\big(\mathbb{R}^{2,1}\big) \otimes \mathbbm{C}({\rm SL}(2, \mathbb{R}))
\overset{\mathrm{morphism}}{\longrightarrow} C({\rm SL}(2, \mathbb{R})) \otimes \mathbbm{C}({\rm SL}(2,
\mathbb{R})),\\
(f \otimes \mathbf{g}) \overset{\mathrm{morphism}}{\longrightarrow} \varphi[(f \otimes \mathbf{g})] \equiv
(\varphi[f] \otimes \mathbf{g}) ,
\end{gather*}
where $f(\vec{p}) \in C(\mathbb{R}^{2,1})$ is {\it lifted} to $\varphi(f(\vec{p})) \equiv f(\mathbf g) \in C({\rm SL}(2,
\mathbb{R}))$.
Looking at the two Hopf algebra structures it is clear that the morphism should satisfy the relation
\begin{gather*}
\varphi(R_{\mathbf{g}}f(\vec{p}))=\text{ad}_{\mathbf{g}} f(\mathbf{h}),
\end{gather*}
which written explicitly reads
\begin{gather*}
\varphi\big(f\big(R\big(\mathbf{g}^{-1}\big)\vec{p}\big)\big)= f\big(\mathbf{g}^{-1}\mathbf{h}\mathbf{g}\big).
\end{gather*}
Clearly the morphism {\it depends} on the choice of group parametrization~\cite{joung} indeed each sets of coordinates
on ${\rm SL}(2, \mathbb{R})$ can be used to map $C(\mathbb{R}^{2,1})$ into $C({\rm SL}(2, \mathbb{R}))$.
Below we will see how group parametrization are associated to dif\/ferent examples of {\it non-commutative plane waves}
(as it has been shown for ${\rm SU}(2)$~\cite{FreidelMajid, oriti,Meljanac:2010et}) and we will link them to the notion of
{\it Weyl map} used in the literature on non-commutative space-times.

\subsection{Non-commutative plane waves and Weyl maps}\label{section4.2}

In order to establish a~link between plane waves and Weyl maps we start by reviewing the dif\/ferent notions of
non-commutative plane waves that are found in the literature on non-commutative spaces in $2+1$ dimensions in connection
with ${\rm SL}(2,\mathbb{R})$ momentum space.

The $\star$-product, that is customarily introduced on $C(\mathbb{R}^{3})$ in the literature on discrete Euclidean
quantum gravity models~\cite{freidel3d, FreidelMajid}, in the Lorentzian case is def\/ined~by
\begin{gather}
e^{\frac{i \kappa}{2}\text{Tr}(X\mathbf{g})} \star e^{\frac{i \kappa}{2}\text{Tr}(X\mathbf{h})} = e^{\frac{i
\kappa}{2}\text{Tr}(X\mathbf{gh})} ,
\label{eq:starp}
\end{gather}
where $\vec{x} \in \mathbb{R}^{2,1}$ are coordinates of $X=x^{\mu}\gamma_{\mu} \in \mathfrak{sl}(2,\mathbb{R})$,
$\mu=1,2,3$.
The group element appearing in the plane wave is taken in the Cartesian parametrization
$\mathbf{g}(\vec{p})=u\mathbbm{1}+ p_{\kappa}^{\mu} \gamma_{\mu} $.
These plane waves can be obtained through a~map~$E$ given~by
\begin{alignat*}{3}
& E: \quad &&  {\rm SL}(2,\mathbb{R}) \to \big\{C\big(\mathbb{R}^{2,1}\big),\star \big\}\equiv C_{\kappa}\big(\mathbb{R}^{2,1}\big), \\
&&& \mathbf{g} \to E_{\mathbf{g}}(\vec{x}) \overset{\mathrm{def}}{=} e^{\frac{i
\kappa}{2}\text{Tr}(X\mathbf{g})}=e^{i \vec{p}\cdot \vec{x}}  , &
\end{alignat*}
where $\{C(\mathbb{R}^{2,1}),\star \}$ is the set of functions on $\mathbb{R}^{2,1}$ equipped with the
$\star$-product~\eqref{eq:starp}.
In this way we have an ordinary plane wave equipped with a~non-commutative star-product determined by the non-abelian
composition rule of the group $e^{i \vec{p}_1 \cdot \vec{x}} \star e^{i \vec{p}_2 \cdot \vec{x}} = e^{i (\vec{p}_1
\oplus \vec{p}_2) \cdot \vec{x}}$, which at leading order in the deformation parameter reads
\begin{gather*}
\vec{p}_1 \oplus \vec{p}_2 = u_{\vec{p}_2}  \vec{p}_1 + u_{\vec{p}_1} \vec{p}_2 +\frac{1}{\kappa}  \vec{p}_1
\wedge_m \vec{p}_2 \simeq \vec{p}_1 + \vec{p}_2 +\frac{1}{\kappa}  \vec{p}_1 \wedge_m \vec{p}_2+
\mathcal{O}\left(\frac{1}{\kappa^2}\right)  ,
\end{gather*}
where $\vec{p}_1 \wedge_m \vec{p}_2 \overset{\mathrm{def}}{=} \text{diag}(-,+,+) \, \vec{p}_1 \wedge
\vec{p}_2$.

Another set of plane waves~\cite{Majid3d} labelled by the group element $\mathbf{g} \in {\rm SL}(2,\mathbb{R})$ is given
in terms of the map
\begin{alignat}{3}
& e:
\quad &&  {\rm SL}(2,\mathbb{R}) \to \hat{C}_{\kappa}\big(\mathbb{R}^{2,1}\big), & \nonumber\\
&&&  \mathbf{g} \to e_{\mathbf{g}}(\hat{x}), &\nonumber\\
&&& \mathbf{g}=e^{k^{\mu} \gamma_{\mu}} \overset{\mathrm{e}}{\to} e_{\mathbf{g}}(\vec{\hat{x}})=e^{i k^{\mu}
\hat{x}_{\mu}}  , &\label{expmap}
\end{alignat}
where $\hat{C}_{\kappa}(\mathbb{R}^{2,1})$ is the non commutative version of $C_{\kappa}(\mathbb{R}^{2,1})$,
i.e.~coordinates on $\mathbb{R}^{2,1}$ equipped with the non trivial Lie bracket
\begin{gather}
[\hat{x}_{\mu}, \hat{x}_{\nu}]=\frac{2i}{\kappa}\epsilon_{\mu \nu}{}^{\rho}\hat{x}_\rho  .
\label{eq;commut}
\end{gather}
Thus $\hat{C}_{\kappa}(\mathbb{R}^{2,1})$ can be seen as the {\it universal enveloping algebra}
$U(\mathfrak{sl}(2,\mathbb{R}))$~\cite{FreidelMajid}.
The $k^{\mu}$ in~\eqref{expmap} are the exponential coordinates introduced before
\begin{gather*}
 u=\cosh |\vec{k}_{\kappa}|,
\qquad
 p_{\mu}=\frac{\sinh |\vec{k}_{\kappa}|}{|\vec{k}_{\kappa}|} k_{\mu} .
\end{gather*}
Notice that the map~$e$ clearly depends on the group parametrization.
Indeed using Cartesian coordinates $p_{\mu}$ we have
\begin{gather*}
\mathbf{g}=e^{k^{\mu}_{\kappa}(\vec{p})\gamma_{\mu}} \overset{\mathrm{e}}{\to}
e_{\mathbf{g}}\big(\vec{\hat{x}}\big)=e^{i \frac{\operatorname{arcsinh} |\vec{p}_{\kappa}|}{|\vec{p}_{\kappa}|} p^{\mu} \hat{x}_{\mu}}
\qquad
\left(=\sqrt{1+|\vec{p}_{\kappa}|^2}\mathbbm{1}+i p^{\mu}\hat{x}_{\mu}\right) .
\end{gather*}
The same could be repeated for other sets of coordinates.
The dif\/ferent plane waves will be linked by a~map~$\phi$ such that
\begin{alignat*}{3}
& \phi: \quad &&  \hat{C}_{\kappa}\big(\mathbb{R}^{2,1}\big) \to C_{\kappa}\big(\mathbb{R}^{2,1}\big), &\\
 &&& e_{\mathbf{g}}\big(\vec{\hat{x}}\big) \to
E_{\mathbf{g}}(\vec{x})=\phi\big(e_{\mathbf{g}} (\vec{\hat{x}} )\big). &
\end{alignat*}
This map is an isomorphism between algebras, in particular considering the product of two non-commuting functions we
have
\begin{gather*}
f_1 (\vec{\hat{x}} )f_2 (\vec{\hat{x}} )
=\phi^{-1}\big(\phi \big(f_1 (\vec{\hat{x}} )\big)\star \phi\big(f_2 (\vec{\hat{x}} )\big)\big)  .
\end{gather*}
Notice that the map $ \Omega \overset{\mathrm{def}}{=} \phi^{-1}: C_{\kappa}(\mathbb{R}^{2,1}) \to
\hat{C}_{\kappa}(\mathbb{R}^{2,1})$ such that
\begin{gather*}
e_{\mathbf{g}} (\vec{\hat{x}} )= \Omega(E_{\mathbf{g}}(\vec{x}))
\overset{\mathrm{def}}{=} \phi^{-1}(E_{\mathbf{g}}(\vec{x}))  ,
\end{gather*}
can be interpreted as a~Weyl map~\cite{Agostini:2002de}.
Indeed the $\star$-product can be written in terms of~$\Omega$ as
\begin{gather}
f_1(\vec{x})\star f_2(\vec{x})=\Omega^{-1}(\Omega(f_1(\vec{x}))  \Omega(f_2(\vec{x})))  ,
\label{eq:pronoc}
\end{gather}
just as in the case of Weyl maps for Lie algebra non-commutative space-times in $3+1$ dimensions as discussed for example
in~\cite{Agostini:2003vg}.
Notice how the notion of $\star$-product def\/ined on commuting function depends on the choice of Weyl map~$\Omega$.

\subsection{(Quantum) group Fourier transform}\label{section4.3}

The non-commutative plane wave $e_{\mathbf{g}}(\vec{\hat{x}})$ is associated to a~notion of {\it quantum group
Fourier transform} which maps functions on a~$n$-dimensional Lie group~$G$ to functions in
$\hat{C}_{\kappa}(\mathbb{R}^n)$~\cite{FreidelMajid}:
\begin{alignat*}{3}
& \mathcal{F}: \quad &&  C(G) \to \hat{C}_{\kappa}(\mathbb{R}^n), &\\
 &&& \tilde{f}(\mathbf{g}) \to F(\vec{\hat{x}})=:\mathcal{F}(\tilde{f}(\mathbf{g}))=\int_G d\mathbf{g}
\tilde{f}(\mathbf{g})  e_{\mathbf{g}}(\vec{\hat{x}}) . &
\end{alignat*}
In the case of ${\rm SL}(2,\mathbb{R})$ associated to the non-commutative plane wave $e_{\mathbf{g}}(\vec{\hat{x}})=e^{i
k^{\mu} \hat{x}_{\mu}}$ we have
\begin{alignat*}{3}
& \mathcal{F}: \quad &&
 C({\rm SL}(2,\mathbb{R})) \to \hat{C}_{\kappa}\big(\mathbb{R}^{2,1}\big), &\\
&&&  \tilde{f}(\mathbf{g}) \to F(\vec{\hat{x}})=:\mathcal{F}(\tilde{f}(\mathbf{g}))=\int_{{\rm SL}(2,\mathbb{R})}
d\vec{k}_{\kappa} \left(\frac{\sinh |\vec{k}_{\kappa}|}{|\vec{k}_{\kappa}|}\right)^2  \tilde{f}(\vec{k})  e^{i
\vec{k}\cdot \vec{\hat{x}}}  ,&
\end{alignat*}
where the Haar measure on the group is expressed in terms of ``exponential'' coordinates $k_{\mu}$ (see Appendix~\ref{appendixD} for
a~review of the Haar measure in the various parametrizations).
This Fourier transform is related to the {\it group Fourier transform} discussed in~\cite{ARZ1, freidel3d, Noui} via the
map~$\phi$ introduced in the previous section.
Indeed the following diagram holds
\begin{gather*}
C(G) \overset{\mathcal{F}}{\to}
\hat{C}_{\kappa}(\mathbb{R}^n)\overset{\phi}{\to} C_{\kappa}(\mathbb{R}^n) .
\end{gather*}
The composition of maps $\mathcal{F} \circ \phi$ is the group Fourier transform
\begin{gather*}
C(G) \overset{\mathcal{F} \circ \phi}{\to} C_{\kappa}(\mathbb{R}^n)  .
\end{gather*}
Notice that the {\it quantization maps} $\mathcal{Q}$ discussed in~\cite{oriti} in our picture coincides with the Weyl
map in~\eqref{eq:pronoc}.
Moreover if we change coordinates on the group and pass from the plane wave in exponential coordinates $\vec{k}$ to
cartesian coordinates $\vec{p}$
\begin{gather*}
\mathbf{g}(\vec{p})=u\mathbbm{1}+p_{\kappa}^{\mu}\gamma_{\mu}=e^{\frac{\operatorname{arcsinh} |\vec{p}_{\kappa}|}{|\vec{p}_{\kappa}|}p_{\kappa}^{\mu}\gamma_{\mu}} \to e^{i \frac{\operatorname{arcsinh}|\vec{p}_{\kappa}|}{|\vec{p}_{\kappa}|}p^{\mu}\hat{x}_{\mu}} ,
\end{gather*}
the quantum group Fourier transform by construction will read
\begin{gather*}
F(\vec{\hat{x}})=:\mathcal{F}(\tilde{f}(\mathbf{g}))=\int_{{\rm SL}(2,\mathbb{R})}
\frac{d\vec{p}_{\kappa}}{\sqrt{1+|\vec{p}_{\kappa}|^2}}
\tilde{f}(\vec{p}) e^{i \frac{\operatorname{arcsinh}|\vec{p}_{\kappa}|}{|\vec{p}_{\kappa}|}p^{\mu}\hat{x}_{\mu}}  .
\end{gather*}
Below we show how to describe these Fourier transforms in a~unif\/ied framework.

\subsection{The master Fourier transform}\label{section4.4}

In this section we exhibit an alternative route to construct non-commutative plane waves and the quantum group Fourier
transform.
This will lead us to an implicit def\/inition of both which for each group parametrization reproduces the corresponding
Fourier transforms found in the literature.
We start by considering the group elements written in the three dif\/ferent sets of coordinates introduced above
\begin{gather*}
 \mathbf{g}(\vec{p})=\sqrt{1+|\vec{p}_{\kappa}|^2}\mathbbm{1}+p^{\mu}_{\kappa}\gamma_{\mu}=e^{\frac{\operatorname{arcsinh}|\vec{p}_{\kappa}|}{|\vec{p}_{\kappa}|}p_{\kappa}^{\mu}\gamma_{\mu}},
\\
 \mathbf{g}(\vec{k})=e^{k^{\mu}_{\kappa}\gamma_{\mu}},
\\
 \mathbf{g}(\vec{\Pi})=e^{\Pi^0_{\kappa}\gamma_0}e^{\Pi_{\kappa}^i \gamma_i}e^{\Pi^0_{\kappa}\gamma_0}.
\end{gather*}
If we take the generators of the algebra $\mathfrak{sl}(2,\mathbb{R})$ as $\gamma_{\mu}
\overset{\mathrm{def}}{=} i \kappa \hat{x}_{\mu}$, we immediately recover the non-commutative plane waves
introduced above, in the cartesian and exponential parametrization, and we also obtain a~new plane wave for the
$\vec{\Pi}$ parametrization
\begin{gather*}
 e_{\mathbf{g}(\vec{p})}(\vec{\hat{x}})=\sqrt{1+|\vec{p}_{\kappa}|^2}\mathbbm{1}+i
p^{\mu}\hat{x}_{\mu}=e^{i\frac{\operatorname{arcsinh}|\vec{p}_{\kappa}|}{|\vec{p}_{\kappa}|}p^{\mu}\hat{x}_{\mu}},
\\
 e_{\mathbf{g}(\vec{k})}(\vec{\hat{x}})=e^{i k^{\mu} \hat{x}_{\mu}},
\\
 e_{\mathbf{g}(\vec{\Pi})}(\vec{\hat{x}})=e^{i \frac{\Pi^0}{2}\hat{x}_0}e^{i\Pi^i \hat{x}_i}e^{i
\frac{\Pi^0}{2}\hat{x}_0},
\end{gather*}
where the coordinates $\hat{x}_{\mu}$ obey the non-trivial Lie bracket~\eqref{eq;commut}.
Notice also that the plane wave in the parametrization $\vec{\Pi}$ is known in the literature on non-commutative spaces
as the ``time symmetrized'' plane wave and has an associated Weyl map~\cite{Agostini:2006zza, Agostini:2003vg}.
The quantum group Fourier transform for the three sets of coordinates reads
\begin{gather*}
 F(\vec{\hat{x}})=\int_{{\rm SL}(2,\mathbb{R})} \frac{d\vec{p}_{\kappa}}{\sqrt{1+|\vec{p}_{\kappa}|^2}}
\tilde{f}(\vec{p})  e^{i \frac{\operatorname{arcsinh}|\vec{p}_{\kappa}|}{|\vec{p}_{\kappa}|}\vec{p}\cdot\vec{\hat{x}}},
\\
 F(\vec{\hat{x}})=\int_{{\rm SL}(2,\mathbb{R})} d\vec{k}_{\kappa} \left(\frac{\sinh
|\vec{k}_{\kappa}|}{|\vec{k}_{\kappa}|}\right)^2 \tilde{f}(\vec{k})  e^{i \vec{k}\cdot \vec{\hat{x}}},
\\
 F(\vec{\hat{x}})=\int_{{\rm SL}(2,\mathbb{R})} d\vec{\Pi}_{\kappa}  \frac{\sinh
2|\underline{\Pi}_{\kappa}|}{2|\underline{\Pi}_{\kappa}|}  \tilde{f}(\vec{\Pi}) e^{i
\frac{\Pi^0}{2}\hat{x}_0}e^{i\Pi^i \hat{x}_i}e^{i \frac{\Pi^0}{2}\hat{x}_0}.
\end{gather*}
Now we can introduce a~formal expression which def\/ines the quantum group Fourier transform in terms of a~generic group
element $\mathbf{g}$.

We def\/ine a~{\it master} plane wave (for ${\rm SL}(2,\mathbb{R})$\footnote{For the case of ${\rm SU}(2)$ we simply
substitute hyperbolic functions with trigonometric functions and Pauli matrices instead of~$\gamma$ matrices.}) given~by
\begin{gather*}
e_{\mathbf{g}}(\vec{\hat{x}})=e^{i\frac{\kappa}{2}\frac{\operatorname{arcsinh} \Delta(\mathbf{g})}{\Delta(\mathbf{g})}
\text{Tr}(\mathbf{g}\gamma_{\mu})\hat{x}^{\mu}} ,
\end{gather*}
where $\Delta(\mathbf{g}) \overset{\mathrm{def}}{=} \frac{1}{2}\sqrt{[\text{Tr}(\mathbf{g})]^2-4}$.
It is easy to show that this def\/inition of plane waves reproduces all the cases above.
We can thus write down an implicit def\/inition of Fourier transform, the {\it master quantum group Fourier transform}
\begin{gather}
\mathfrak{F}(\vec{\hat{x}})=:\mathcal{F}(\tilde{f}(\mathbf{g})) \overset{\mathrm{def}}{=} \int_G
d\mu(\mathbf{g})   \tilde{f}(\mathbf{g})   e^{i\mathcal{K}_{\mu}(\mathbf{g})\hat{x}^{\mu}} ,
\label{eq:mastertrasf}
\end{gather}
where
\begin{gather*}
 \mathcal{K}_{\mu}(\mathbf{g}) \overset{\mathrm{def}}{=} \frac{\kappa}{2}\frac{\operatorname{arcsinh}
\Delta(\mathbf{g})}{\Delta(\mathbf{g})} \text{Tr}(\mathbf{g}\gamma_{\mu}),
\qquad
 \Delta(\mathbf{g}) \overset{\mathrm{def}}{=} \frac{1}{2}\sqrt{[\text{Tr}(\mathbf{g})]^2-4}  .
\end{gather*}
Equation~\eqref{eq:mastertrasf} provides a~formal def\/inition of (quantum) group Fourier transform which reproduces the known
Fourier transforms appeared in the literature when a~quantization/Weyl map is specif\/ied\footnote{The implicit Fourier
transform~\eqref{eq:mastertrasf} is analogous to the general form of ``non-commutative'' Fourier transform discussed in
Section~IV of~\cite{oriti}.}.
We conclude by stressing that the choice of quantization (Weyl) map determines uniquely the form of the plane waves
entering the (non-commutative) Fourier transform and is equivalent to a~choice of coordinates on the Lie group momentum
space~\cite{oriti}.

\section{Conclusions}\label{section5}

In this work we of\/fered a~quick, but we hope comprehensive, journey through the notions of deformed symmetry and
non-commutativity which are encountered in the study of a~relativistic point particle coupled to three-dimensional
gravity.
In doing so we reviewed and clarif\/ied various notions that appear scattered in the literature but also provided some
valuable new insights on these models.

On one side we showed for the f\/irst time, introducing the new set of {\it Euler coordinates}, how the framework of
doubly or deformed special relativity is realized in the context of ${\rm SL}(2,\mathbb{R})$ momentum space.
In particular we showed that the model considered introduces a~notion of maximal energy which is compatible with
a~deformed action of boosts on momenta living on a~group manifold.
From a~more abstract point of view we connected several important notions which have appeared in the past in the
literature on non-commutative spaces with those recently studied within the community working on {\it group field
theory}~\cite{Baratin:2010nn, Baratin:2010wi}.
Specif\/ically we clarif\/ied the status of the dif\/ferent notions of (quantum) group Fourier transforms appearing in the
literature and we were able to formulate an implicit {\it master Fourier transform} which, for each choice of
star-product or Weyl map, reproduces the various group Fourier transforms proposed in the literature.
Particularly signif\/icant is the explicit connection we established between dif\/ferent notions of group Fourier
transforms, Weyl maps and choices of coordinates on the momentum group manifold.
Such connection provides further evidence that models resulting from dif\/ferent quantizations/Weyl maps correspond to
dif\/ferent physical scenarios~\cite{Arzano:2010jw}.
Moreover our f\/indings bridge an important gap between the two f\/ields and we believe this work will provide a~useful
reference for those seeking cross-fertilization between the techniques used in the research areas of non-commutative
f\/ield theory and group f\/ield theory.

\appendix

\section[${\rm SL}(2,\mathbb{R})$]{$\boldsymbol{{\rm SL}(2,\mathbb{R})}$}\label{appendixA}

  In this section we want to characterize the Lie group ${\rm SL}(2,\mathbb{R})$, in particular we will recall
its basic general properties and its description as a~manifold.

\subsection{Basic general properties}\label{sectionA.1}

The group ${\rm SL}(2,\mathbb{R})$ is represented by the set of all $2 \times 2$ matrices
\begin{gather*}
\mathcal{M}=
\begin{bmatrix}
M_{11} &  M_{12}
\\
M_{21} &  M_{22}
\end{bmatrix}
,
\end{gather*}
in which $M_{ij} \in \mathbb{R}$ and $\det \mathcal{M}= M_{11} M_{22}- M_{21} M_{12}=1$.
The number of free parameters, considering the determinant constraint, is $n=4-1 =3$.
A~way to represents a~group element~${\mathbf g}$ as a~matrix of ${\rm SL}(2,\mathbb{R})$ is using the
expansion in terms of the unit and the~$\gamma$ matrices
\begin{gather*}
\gamma_0=
\begin{bmatrix}
0 &  1
\\
-1  & 0
\end{bmatrix},
\qquad
\gamma_1=
\begin{bmatrix}
0 &  1
\\
1 &  0
\end{bmatrix},
\qquad
\gamma_2=
\begin{bmatrix}
1  & 0
\\
0 &  -1
\end{bmatrix}
\qquad
(\text{such that}
\quad
[\gamma_{\mu},\gamma_{\nu}]=f_{\mu \nu}^{\rho}\gamma_{\rho}) ,
\end{gather*}
namely
\begin{gather*}
{\bf g}=u\mathbbm{1}+\xi^{\mu} \gamma_{\mu} ,
\end{gather*}
where~$u$, $\xi^0$, $\xi^1$, $\xi^2$ are real parameters.
This expression can be explicitly rewritten as a~matrix
\begin{gather*}
{\bf g}=
\begin{bmatrix}
u+\xi^2  & \xi^1+\xi^0
\\
\xi^1-\xi^0  & u-\xi^2
\end{bmatrix}
,
\end{gather*}
whose determinant constraint is $\det {\mathbf g}=u^2-(\xi^1)^2-(\xi^2)^2+(\xi^0)^2=1$.
This equation shows that the~$u$ parameter can be expressed as a~function of the other parameters:
$u^2=1-(\xi^0)^2+(\xi^1)^2+(\xi^2)^2$.
At this point, we make another step, namely characterize geometrically the group ${\rm SL}(2,\mathbb{R})$.

\subsection[${\rm SL}(2,\mathbb{R})$ as a~manifold]{$\boldsymbol{{\rm SL}(2,\mathbb{R})}$ as a~manifold} \label{appendixA.2}

\begin{figure}[htbp] \centering
\includegraphics[width=7cm]{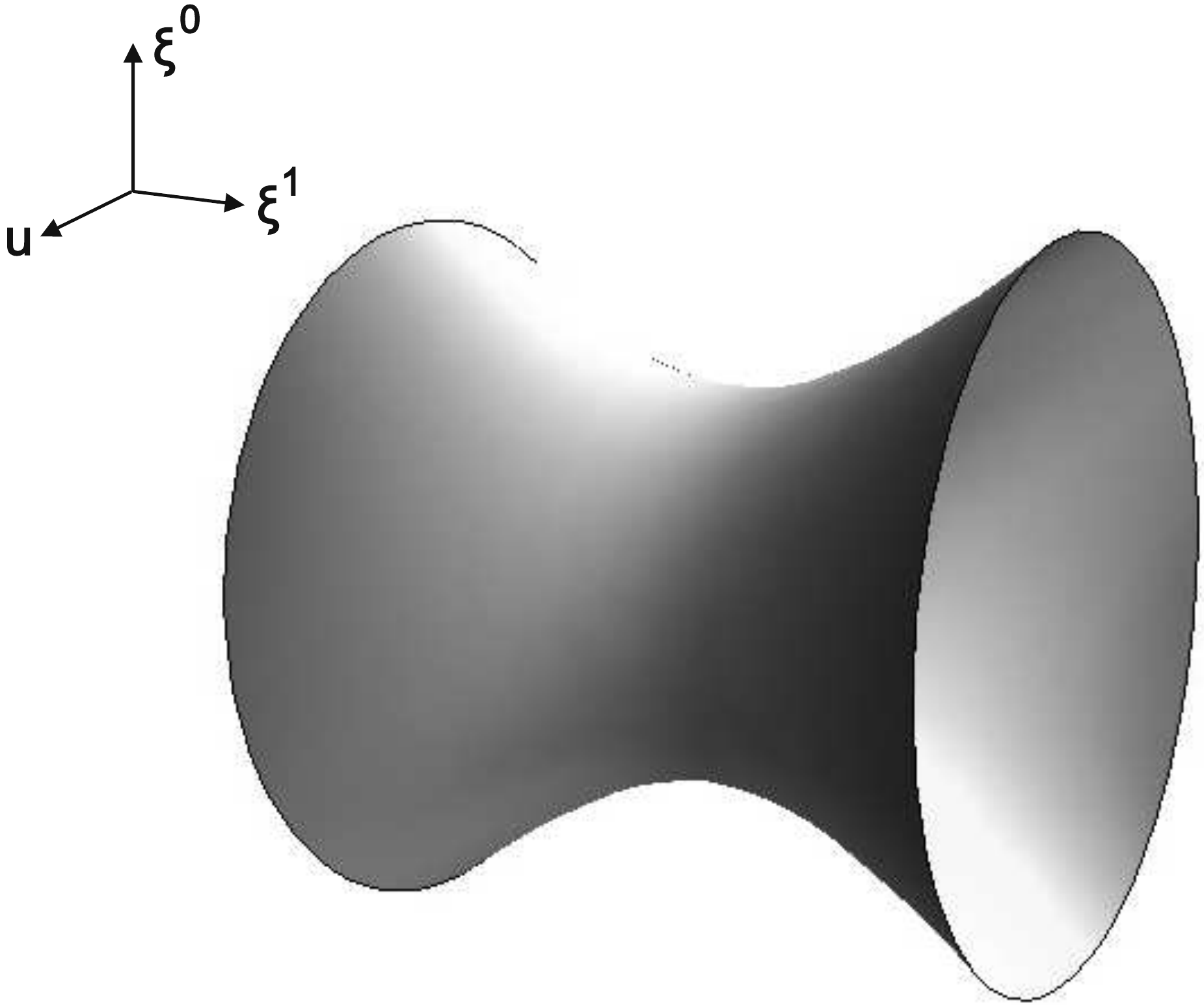}

\caption{The group manifold ${\rm
SL}(2,\mathbb{R})$ embedded in $\mathbb{R}^{2,2}$.
The coordinates are $(\xi^{\mu},u)$ with $\xi^2$ suppressed.
This space is known as {\it Anti de Sitter} (AdS).}\label{ADS}
\end{figure}

  We start this section considering the determinant constraint just written
\begin{gather*}
\big(\xi^0\big)^2-\big(\xi^1\big)^2-\big(\xi^2\big)^2+u^2=1 .
\end{gather*}
This equation def\/ines an hyperboloid embedded in $\mathbb{R}^{2,2}$ (signature $(+,-,-,+)$, see Fig.~\ref{ADS}).

\begin{figure}[htbp] \centering
\includegraphics[width=5cm]{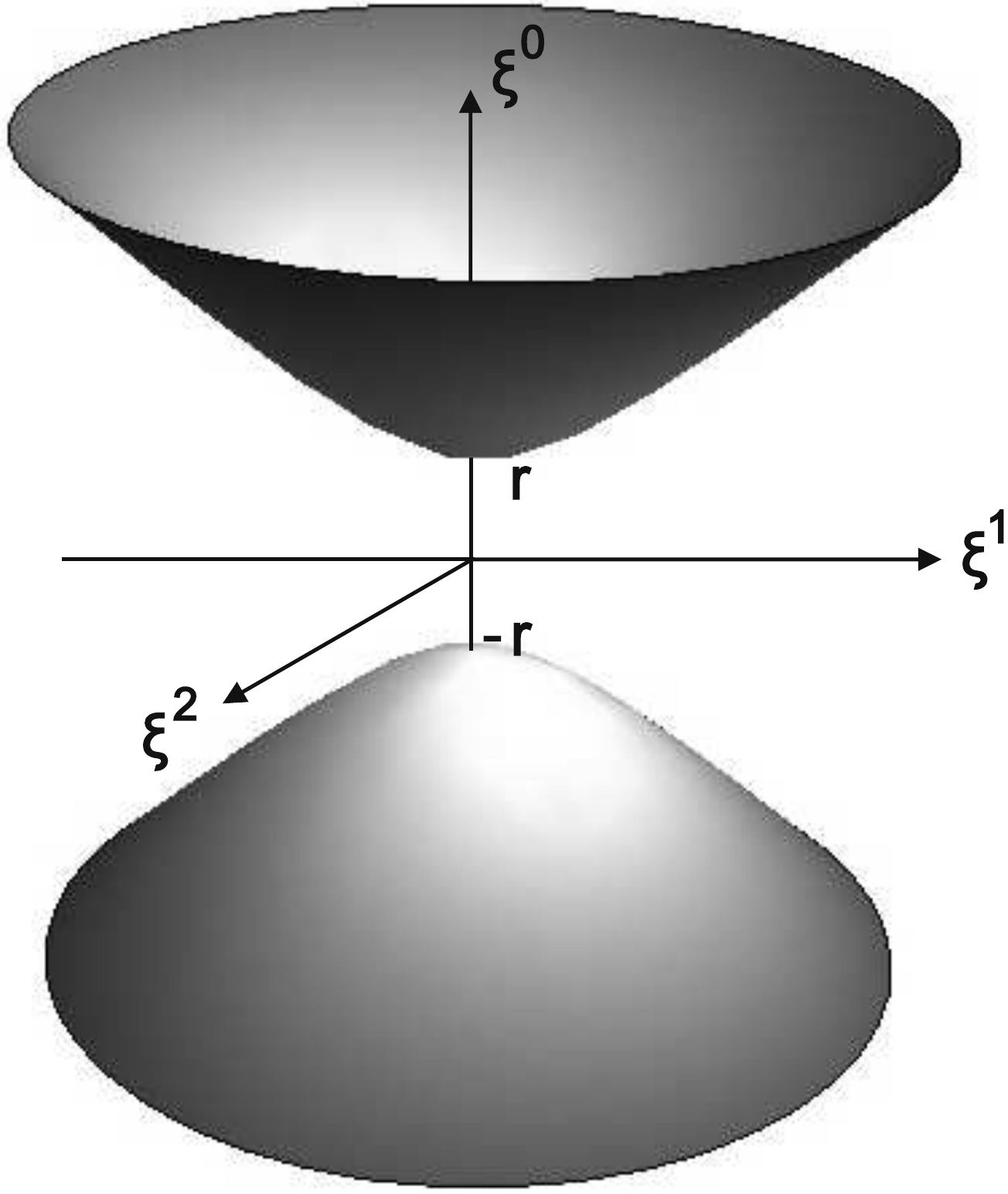}

\caption{Space of parameters $\xi^{\mu}$
($\mu=0,1,2$) for $|u|<1$.
This elliptic spatial geometry is known as Lobacevskij space.}\label{tsheets}
\end{figure}

 Now it is clear that, for dif\/ferent values of the parameter~$u$, we have dif\/ferent geometries for the space of
the free parameters~$\xi^{\mu}$.
In particular if we write
\begin{gather}
\big(\xi^0\big)^2-\big(\xi^1\big)^2-\big(\xi^2\big)^2=1-u^2,
\label{eq:detconn}
\end{gather}
for $1-u^2>0 \Leftrightarrow (1-u)(1+u)>0$ (namely $|u|<1$), we can rewrite the equation~\eqref{eq:detconn}, def\/ining
the quantity $r^2\doteq 1-u^2 > 0$, as
\begin{gather*}
\big(\xi^0\big)^2-\big(\xi^1\big)^2-\big(\xi^2\big)^2= r^2,
\end{gather*}
which is the equation of an elliptic hyperboloid (see Fig.~\ref{tsheets}).

\begin{figure}[ht!] \centering
\includegraphics[width=5cm]{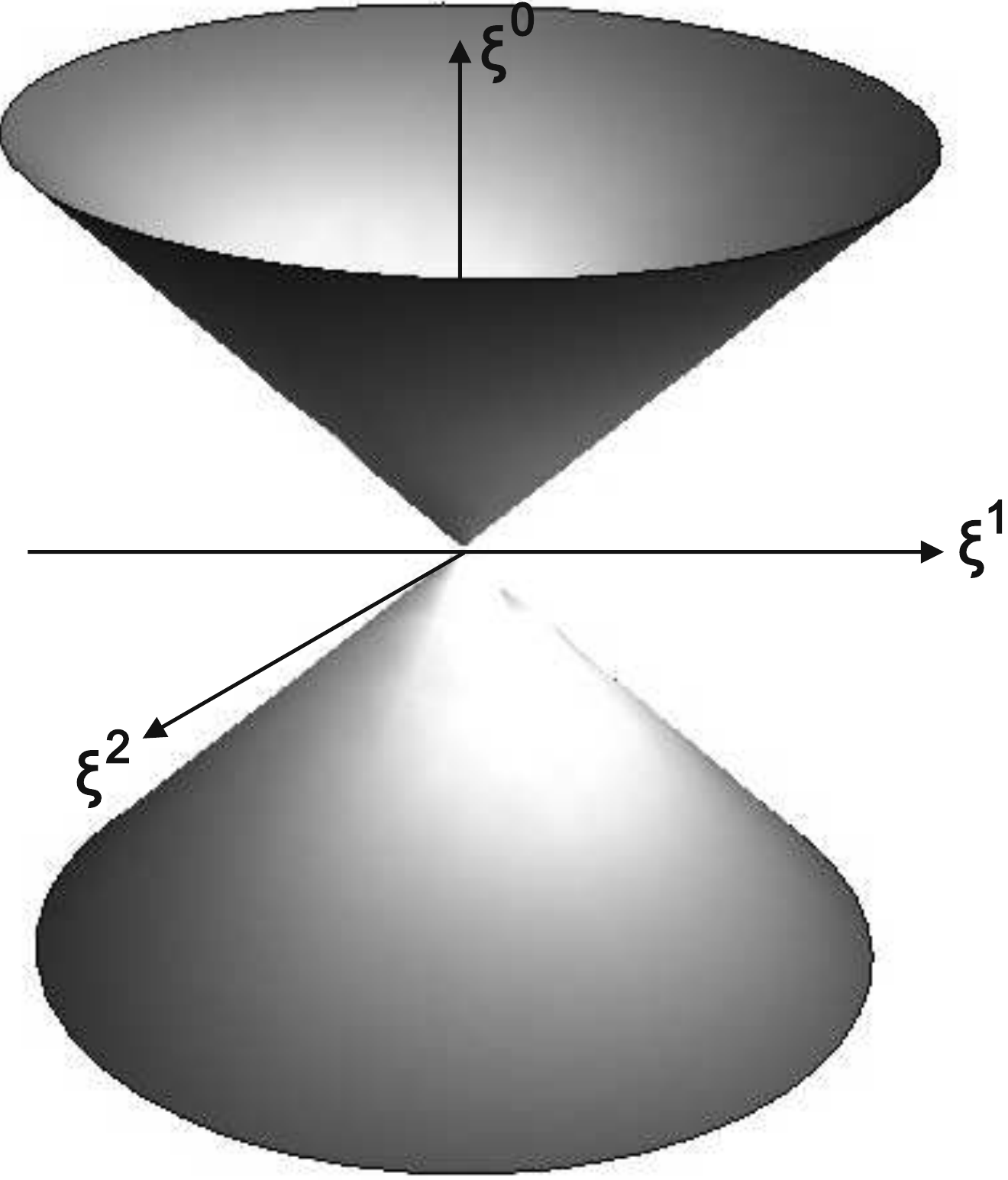}

\caption{Space of parameters $\xi^{\mu}$ ($\mu=0,1,2$) for $|u|=1$.}
\label{conem}
\end{figure}

 Afterwards, if we take the quantity $1-u^2=0$ (namely $|u|=1$), the equation becomes{\samepage
\begin{gather*}
\big(\xi^0\big)^2-\big(\xi^1\big)^2-\big(\xi^2\big)^2=0,
\end{gather*}
which is the equation of a~cone (see Fig.~\ref{conem}).}

\begin{figure}[ht!] \centering
\includegraphics[width=5cm]{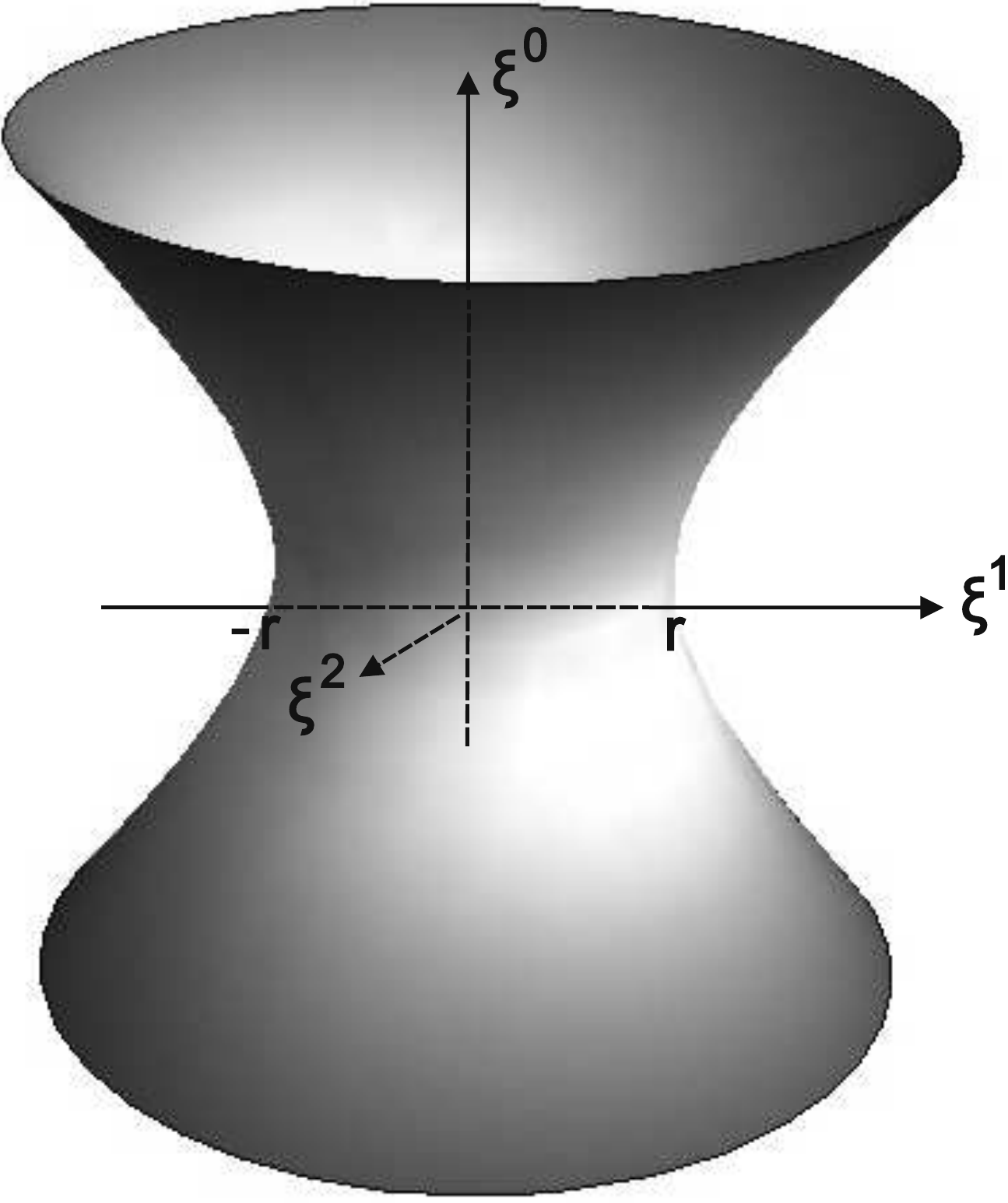}

\caption{Space of parameters $\xi^{\mu}$ ($\mu=0,1,2$) for $|u|>1$.}\label{des}
\end{figure}

Finally, if we take $1-u^2<0$ (namely $|u|>1$), the determinant reads{\samepage
\begin{gather*}
\big(\xi^0\big)^2-\big(\xi^1\big)^2-\big(\xi^2\big)^2=-r^2 \ \Rightarrow \  -\big(\xi^0\big)^2+\big(\xi^1\big)^2+\big(\xi^2\big)^2=r^2,
\end{gather*}
which is the equation of an hyperbolic hyperboloid (see Fig.~\ref{des}).}

\begin{figure}[htbp] \centering \includegraphics[width=10cm]{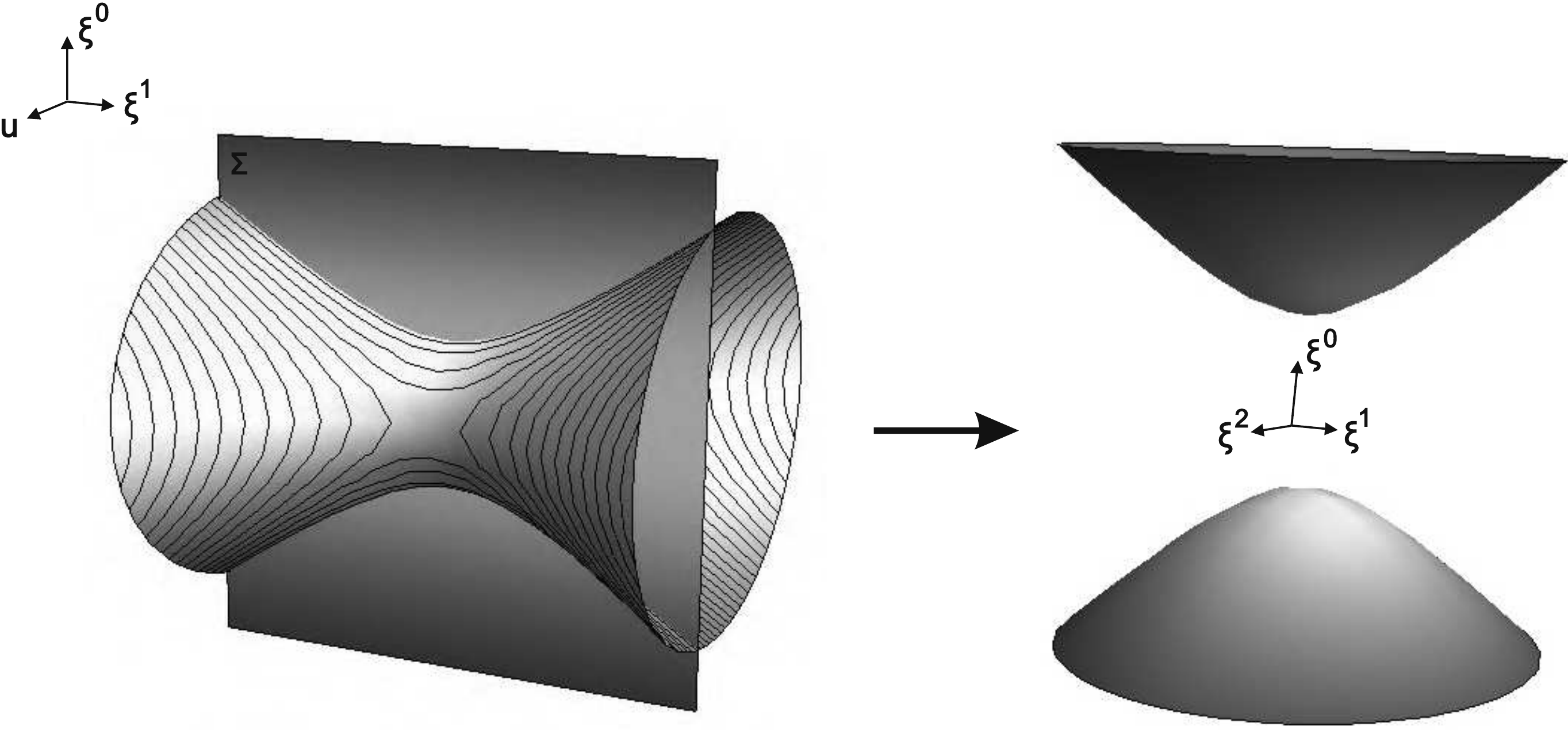}

\caption{The left part of the f\/igure shows
the intersection of the~$\Sigma$-plane, which represents a~surface with $u=\text{const}$, with the AdS-manifold.
The value of the constant is such that $|u|<1$ and, as seen before, this implies a~Lobacevskij geometry, obtained
rotating the intersection of the~$\Sigma$-plane, for the parameter space (right part of the f\/igure).}
\label{manint}
\end{figure}

These are all the geometries of the parameter space of the group ${\rm SL}(2,\mathbb{R})$.
In particular, we must imagine that for a~f\/ixed value of the parameter~$u$, also known as embedding parameter, we
generate one of these geometries.
We can clarify our ideas observing Fig.~\ref{manint}.

\section{Euler coordinates expansion}\label{appendixB}

Starting by the following expression for the group element
\begin{gather*}
{\mathbf g}= e^{\frac{1}{2}(\rho_{\kappa} + \phi)\gamma_0}e^{\chi_{\kappa} \gamma_1}e^{\frac{1}{2}(\rho_{\kappa} -
\phi)\gamma_0}  ,
\end{gather*}
in terms of the dimensionful parameters $\rho_{\kappa}$ and $\chi_{\kappa}$, we can separate the exponentials as follows
\begin{gather}
{\mathbf g}= e^{\frac{1}{2}\rho_{\kappa}\gamma_0} e^{\frac{1}{2}\phi\gamma_0} e^{\chi_{\kappa}
\gamma_1} e^{-\frac{1}{2}\phi\gamma_0}  e^{\frac{1}{2}\rho_{\kappa}\gamma_0} ,
\end{gather}
this is because $\gamma_0$, obviously, commutes with itself and then we can divide the exponential in two parts.
Now, we focus on the three central exponentials.

Rewriting the latter in matrix form we have
\begin{gather*}
e^{\frac{1}{2}\phi\gamma_0} e^{\chi_{\kappa} \gamma_1} e^{-\frac{1}{2}\phi\gamma_0}
 =\left[\cos\frac{\phi}{2} \mathbbm{1}+\sin \frac{\phi}{2} \gamma_0\right]\left[\cosh\chi_{\kappa} \mathbbm{1}+\sinh
\chi_{\kappa} \gamma_1\right]\left[\cos\frac{\phi}{2} \mathbbm{1}-\sin \frac{\phi}{2} \gamma_0\right]
\\
\hphantom{e^{\frac{1}{2}\phi\gamma_0} e^{\chi_{\kappa} \gamma_1} e^{-\frac{1}{2}\phi\gamma_0} }{}
 =\cosh \chi_{\kappa}  \mathbbm{1}+\left[\cos^2 \frac{\phi}{2}-\sin^2 \frac{\phi}{2}\right]\sinh \chi_{\kappa}
\gamma_1 +2\cos \frac{\phi}{2}\sin \frac{\phi}{2}\sinh \chi_{\kappa}   \gamma_2
\\
\hphantom{e^{\frac{1}{2}\phi\gamma_0} e^{\chi_{\kappa} \gamma_1} e^{-\frac{1}{2}\phi\gamma_0} }{}
 =\cosh \chi_{\kappa} \mathbbm{1}+\cos\phi \sinh \chi_{\kappa} \gamma_1+\sin\phi \sinh \chi_{\kappa}   \gamma_2.
\end{gather*}
At this point, considering that we def\/ined our spatial momenta as
\begin{gather*}
 \Pi^1 \doteq \chi\cos\phi=|\underline{\Pi}|\cos\phi,
\qquad
\Pi^2 \doteq \chi\sin\phi=|\underline{\Pi}|\sin\phi,
\end{gather*}
substituting these expressions we obtain
\begin{gather*}
e^{\frac{1}{2}\phi\gamma_0} e^{\chi_{\kappa} \gamma_1} e^{-\frac{1}{2}\phi\gamma_0}
=\cosh \chi_{\kappa}
 \mathbbm{1}+\cos\phi \sinh \chi_{\kappa}  \gamma_1+\sin\phi \sinh \chi_{\kappa}   \gamma_2
\\
\hphantom{e^{\frac{1}{2}\phi\gamma_0} e^{\chi_{\kappa} \gamma_1} e^{-\frac{1}{2}\phi\gamma_0} }{}
 =\cosh |\underline{\Pi}_{\kappa}|  \mathbbm{1}+\frac{\sinh |\underline{\Pi}_{\kappa}|}{|\underline{\Pi}|}\big[\Pi^1
\gamma_1+\Pi^2\gamma_2\big]
\\
\hphantom{e^{\frac{1}{2}\phi\gamma_0} e^{\chi_{\kappa} \gamma_1} e^{-\frac{1}{2}\phi\gamma_0} }{}
 =\cosh |\underline{\Pi}_{\kappa}|  \mathbbm{1}+\frac{\sinh
|\underline{\Pi}_{\kappa}|}{|\underline{\Pi}_{\kappa}|}\big[\Pi_{\kappa}^1 \gamma_1+\Pi_{\kappa}^2\gamma_2\big]
 =e^{\Pi_{\kappa}^1 \gamma_1+\Pi_{\kappa}^2\gamma_2},
\end{gather*}
where the last equality can be easily obtained using the properties of the~$\gamma$ matrices in the resummation of the
exponential.
In conclusion, substituting also the Euler angle~$\rho$ with our momentum~$\Pi^0$ and putting together the exponentials,
we obtain the expression~\eqref{eq:piform}.

\section{Change of coordinates between dif\/ferent parametrizations}\label{appendixC}

Let us start by writing all together the set of coordinates on the manifold we considered
\begin{gather}
 u=\sqrt{1+|\vec{p}_{\kappa}|^2}=\cosh|\vec{k}_{\kappa}|=\cosh|\underline{\Pi}_{\kappa}|\cos\Pi_{\kappa}^0,\nonumber\\
\xi^0=p_{\kappa}^0= \frac{\sinh|\vec{k}_{\kappa}|}{|\vec{k}_{\kappa}|}k^0_{\kappa}
=\cosh|\underline{\Pi}_{\kappa}|\sin\Pi_{\kappa}^0,\nonumber\\
\xi^1= p_{\kappa}^1=
\frac{\sinh|\vec{k}_{\kappa}|}{|\vec{k}_{\kappa}|}k^1_{\kappa}=\frac{\sinh|\underline{\Pi}_{\kappa}|}{|\underline{\Pi}_{\kappa}|}\Pi_{\kappa}^1,
\nonumber\\
\xi^2=
p_{\kappa}^2=\frac{\sinh|\vec{k}_{\kappa}|}{|\vec{k}_{\kappa}|}k^2_{\kappa}
=\frac{\sinh|\underline{\Pi}_{\kappa}|}{|\underline{\Pi}_{\kappa}|}\Pi_{\kappa}^2,
\label{eq:change acoord}
\end{gather}

Now, starting by the system~\eqref{eq:change acoord}, we can f\/ind the transformation laws from a~set of
coordinates to another.

\subsection[$\vec{p}$ vs $\vec{k}$]{$\boldsymbol{\vec{p}}$ vs $\boldsymbol{\vec{k}}$}
\vspace{-2mm}

\begin{gather*}
\begin{cases}
p^0= \dfrac{\sinh|\vec{k}_{\kappa}|}{|\vec{k}_{\kappa}|} k^0,
\vspace{1mm}\\
p^1= \dfrac{\sinh|\vec{k}_{\kappa}|}{|\vec{k}_{\kappa}|}k^1,
\vspace{1mm}\\
p^2=\dfrac{\sinh|\vec{k}_{\kappa}|}{|\vec{k}_{\kappa}|}k^2,
\end{cases}
\qquad
\begin{cases} k^0= \dfrac{\operatorname{arcsinh}|\vec{p}_{\kappa}|}{|\vec{p}_{\kappa}|} p^0,
\vspace{1mm}\\
k^1= \dfrac{\operatorname{arcsinh}|\vec{p}_{\kappa}|}{|\vec{p}_{\kappa}|} p^1,
\vspace{1mm}\\
k^2= \dfrac{\operatorname{arcsinh}|\vec{p}_{\kappa}|}{|\vec{p}_{\kappa}|} p^2 .
\end{cases}
\end{gather*}

\subsection[$\vec{p}$ vs $\vec{\Pi}$]{$\boldsymbol{\vec{p}}$ vs $\boldsymbol{\vec{\Pi}}$}
\vspace{-2mm}

\begin{gather*}
\begin{cases} p^0=\kappa \cosh|\underline{\Pi}_{\kappa}|\sin\Pi_{\kappa}^0,
\vspace{1mm}\\
p^1= \dfrac{\sinh|\underline{\Pi}_{\kappa}|}{|\underline{\Pi}_{\kappa}|}\Pi^1,
\vspace{1mm}\\
p^2= \dfrac{\sinh|\underline{\Pi}_{\kappa}|}{|\underline{\Pi}_{\kappa}|}\Pi^2,
\end{cases}
\qquad
\begin{cases}
\Pi^0=\kappa \arctan \left(\dfrac{p_{\kappa}^0}{\sqrt{1+|\vec{p}_{\kappa}|^2}}\right),
\vspace{1mm}\\
\Pi^1= \dfrac{\operatorname{arcsinh}|\underline{p}_{\kappa}|}{|\underline{p}_{\kappa}|}  p^1,
\vspace{1mm}\\
\Pi^2= \dfrac{\operatorname{arcsinh}|\underline{p}_{\kappa}|}{|\underline{p}_{\kappa}|} p^2.
\end{cases}
\end{gather*}

\subsection[$\vec{k}$ vs $\vec{\Pi}$]{$\boldsymbol{\vec{k}}$ vs $\boldsymbol{\vec{\Pi}}$}
\vspace{-2mm}

\begin{gather*}
\begin{cases}
\Pi^0= \kappa \arctan\left(\dfrac{\tanh|\vec{k}_{\kappa}|}{|\vec{k}_{\kappa}|}k_{\kappa}^0\right),
\vspace{1mm}\\
\Pi^1= \dfrac{\operatorname{arcsinh} \left(\frac{\sinh|\vec{k}_{\kappa}|}{|\vec{k}_{\kappa}|}|\underline{k}_{\kappa}|\right)}{|\underline{k}_{\kappa}|}
k^1,
\vspace{1mm}\\
\Pi^2= \dfrac{\operatorname{arcsinh}\left(\frac{\sinh|\vec{k}_{\kappa}|}{|\vec{k}_{\kappa}|}|\underline{k}_{\kappa}|\right)}{|\underline{k}_{\kappa}|}  k^2,
\end{cases} \!\quad
\begin{cases} k^0=\dfrac{\operatorname{arcsinh} \sqrt{\cosh^2|\underline{\Pi}_{\kappa}| \cos^2
\Pi^0_{\kappa}-1}}{\sqrt{\cosh^2|\underline{\Pi}_{\kappa}| \cos^2 \Pi^0_{\kappa}-1}} \\
\hphantom{k^0=}{}\times \kappa\cosh
|\underline{\Pi}_{\kappa}|\cos\Pi_{\kappa}^0,
\vspace{1mm}\\
k^1=\dfrac{\operatorname{arcsinh} \sqrt{\cosh^2|\underline{\Pi}_{\kappa}| \cos^2
\Pi^0_{\kappa}-1}}{\sqrt{\cosh^2|\underline{\Pi}_{\kappa}| \cos^2 \Pi^0_{\kappa}-1}} \dfrac{\sinh
|\underline{\Pi}_{\kappa}|}{|\underline{\Pi}_{\kappa}|} \Pi^1,
\vspace{1mm}\\
k^2=\dfrac{\operatorname{arcsinh} \sqrt{\cosh^2|\underline{\Pi}_{\kappa}| \cos^2
\Pi^0_{\kappa}-1}}{\sqrt{\cosh^2|\underline{\Pi}_{\kappa}| \cos^2 \Pi^0_{\kappa}-1}} \dfrac{\sinh
|\underline{\Pi}_{\kappa}|}{|\underline{\Pi}_{\kappa}|} \Pi^2 .
\end{cases}\!\!\!\!\!\!\!
\end{gather*}

\section{The Haar measure}\label{appendixD}

Following~\cite{fuchs, MATS}, we can write the Haar measure on ${\rm SL}(2,\mathbb{R})$ in terms of the
coordinates which parametrize the group element ${\mathbf g}=u\mathbbm{1}+\xi^{\mu}\gamma_{\mu}$ as
\begin{gather}
d\mu_{{\mathbf g}(\vec{\xi})}  \doteq du d\vec{\xi}  \delta(\det {\mathbf
g}-1)=\frac{d\vec{\xi}}{\sqrt{1+|\vec{\xi}|^2}},
\label{opopop}
\end{gather}
where, as always, $|\vec{\xi}|^2=-(\xi^0)^2+|\underline{\xi}|^2$.
In this way we can characterize the Haar measure on ${\rm SL}(2,\mathbb{R})$ for every parametrization.

\subsection{Haar measure in Cartesian parametrization}\label{appendixD.1}
Using the $\vec{p}$-parametrization we have
\begin{gather*}
 u=\sqrt{1+|\vec{p}_{\kappa}|^2}, \qquad \xi^0=p_{\kappa}^0, \qquad \xi^1= p^1_{\kappa}, \qquad \xi^2= p^2_{\kappa},
\end{gather*}
is immediate to write the Haar measure on the group; in fact, substituting the parameters $\xi^{\mu}$ in the
formula~\eqref{opopop}, we obtain
\begin{gather*}
d\mu_{{\mathbf g}(\vec{p})}=\frac{d\vec{p}_{\kappa}}{\sqrt{1+|\vec{p}_{\kappa}|^2}}.
\end{gather*}

\subsection{Haar measure in exponential parametrization}\label{appendixD.2}

Using the $\vec{k}$-parametrization the matter is more complicated.
This is because these coordinates appear as a~combination of hyperbolic functions
\begin{gather*}
 u=\cosh|\vec{k}_{\kappa}|, \qquad
\xi^0= \frac{\sinh|\vec{k}_{\kappa}|}{|\vec{k}_{\kappa}|}k^0_{\kappa}, \qquad
\xi^1= \frac{\sinh|\vec{k}_{\kappa}|}{|\vec{k}_{\kappa}|}k^1_{\kappa}, \qquad
\xi^2= \frac{\sinh|\vec{k}_{\kappa}|}{|\vec{k}_{\kappa}|}k^2_{\kappa},
\end{gather*}
and thus the form of the Haar measure~\eqref{opopop} is more dif\/f\/icult to obtain.
In particular, we have to calculate the determinant of the Jacobian matrix~$J$ related to the transformation $\vec{\xi}
\to \vec{k}$.
Performing the calculation, we obtain
\begin{gather*}
\det J=
\begin{vmatrix}
\dfrac{\partial \xi^0(\vec{k})}{\partial k^0} & \dfrac{\partial \xi^0(\vec{k})}{\partial k^1} & \dfrac{\partial
\xi^0(\vec{k})}{\partial k^2}
\vspace{1mm}\\
\dfrac{\partial \xi^1(\vec{k})}{\partial k^0} & \dfrac{\partial \xi^1(\vec{k})}{\partial k^1} &  \dfrac{\partial
\xi^1(\vec{k})}{\partial k^2}
\vspace{1mm}\\
\dfrac{\partial \xi^2(\vec{k})}{\partial k^0} &  \dfrac{\partial \xi^2(\vec{k})}{\partial k^1} & \dfrac{\partial
\xi^2(\vec{k})}{\partial k^2}
\end{vmatrix}
=\frac{\cosh |\vec{k}_{\kappa}|\sinh^2 |\vec{k}_{\kappa}|}{\kappa^3|\vec{k}_{\kappa}|^2},
\end{gather*}
which, given equation~\eqref{opopop}, allows us to write the Haar measure as
\begin{gather*}
d \mu_{{\mathbf g}({\vec{k}})}=\frac{\sinh^2 |\vec{k}_{\kappa}|}{|\vec{k}_{\kappa}|^2} d\vec{k}_{\kappa}.
\end{gather*}

\subsection{Haar measure in Euler parametrization}\label{appendixD.3}

The Haar measure in the $\vec{\Pi}$-parametrization can be found in the same way as the previous ones.
In particular the new parametrization
\begin{gather*}
u=\cosh|\underline{\Pi}_{\kappa}|\cos\Pi_{\kappa}^0,
\qquad\!\!
\xi^0= \cosh|\underline{\Pi}_{\kappa}|\sin\Pi_{\kappa}^0,
\qquad\!\!
\xi^1= \frac{\sinh|\underline{\Pi}_{\kappa}|}{|\underline{\Pi}_{\kappa}|}\Pi_{\kappa}^1,
\qquad\!\!
\xi^2= \frac{\sinh|\underline{\Pi}_{\kappa}|}{|\underline{\Pi}_{\kappa}|}\Pi_{\kappa}^2,\!
\end{gather*}
leads to the following Jacobian determinant
\begin{gather*}
\det J=
\begin{vmatrix}
\dfrac{\partial \xi^0(\vec{\Pi})}{\partial \Pi^0} & \dfrac{\partial \xi^0(\vec{\Pi})}{\partial \Pi^1} & \dfrac{\partial
\xi^0(\vec{\Pi})}{\partial \Pi^2}
\vspace{1mm}\\
\dfrac{\partial \xi^1(\vec{\Pi})}{\partial \Pi^0} & \dfrac{\partial \xi^1(\vec{\Pi})}{\partial \Pi^1}  & \dfrac{\partial
\xi^1(\vec{\Pi})}{\partial \Pi^2}
\vspace{1mm}\\
\dfrac{\partial \xi^2(\vec{\Pi})}{\partial \Pi^0} &  \dfrac{\partial \xi^2(\vec{\Pi})}{\partial \Pi^1} & \dfrac{\partial
\xi^2(\vec{\Pi})}{\partial \Pi^2}
\end{vmatrix}
=\frac{\cosh^2|\underline{\Pi}_{\kappa}|\cos\Pi_{\kappa}^0\sinh
|\underline{\Pi}_{\kappa}|}{\kappa^3|\underline{\Pi}_{\kappa}|},
\end{gather*}
which implies that the Haar measure in this parametrization is given~by
\begin{gather*}
d \mu_{{\mathbf g}({\vec{\Pi}})}=\frac{\cosh^2|\underline{\Pi}_{\kappa}|\cos\Pi_{\kappa}^0\sinh
|\underline{\Pi}_{\kappa}| d\vec{\Pi}}{\kappa^3 |\underline{\Pi}_{\kappa}|
\cosh|\underline{\Pi}_{\kappa}|\cos\Pi_{\kappa}^0},
\end{gather*}
namely
\begin{gather*}
d \mu_{{\mathbf g}({\vec{\Pi}})}=\frac{\cosh|\underline{\Pi}_{\kappa}|\sinh
|\underline{\Pi}_{\kappa}|}{|\underline{\Pi}_{\kappa}|}d\vec{\Pi}_{\kappa}
=\frac{\sinh 2|\underline{\Pi}_{\kappa}|}{2 |\underline{\Pi}_{\kappa}|}d\vec{\Pi}_{\kappa}.
\end{gather*}

\subsection*{Acknowledgments}

We would like to thank the anonymous referees for the insightful comments which helped us improve the paper.
MA work is supported by a~Marie Curie Career Integration Grant within the 7th European Community Framework Programme and
in part by a~grant from the John Templeton Foundation.

\pdfbookmark[1]{References}{ref}

\LastPageEnding

\end{document}